\documentclass[3p,times]{elsarticle}

\biboptions{numbers,sort&compress}

\usepackage{graphicx}
\RequirePackage[dvipsnames]{xcolor}

\usepackage{subcaption}

\usepackage{amsmath} % assumes amsmath package installed
\usepackage{amssymb}  % assumes amsmath package installed
\usepackage{amsthm}
\theoremstyle{definition}
\newtheorem{definition}{Definition}[section]
\newtheorem{remark}{Remark}[section]
\newtheorem{theorem}{Theorem}[section]

\usepackage{algorithm}
\usepackage{algpseudocode}

\newcommand{\im}{\textrm{i}}
\newcommand{\R}{\mathbb{R}}
\newcommand{\K}{\mathcal{K}}
\newcommand{\bk}{\mathbf{k}}
\newcommand{\E}{\mathcal{E}}
\newcommand{\cut}{\text{cut}}
\newcommand{\tr}{\text{tr}}
\newcommand\red[1]{{\color{red}#1}}
\newcommand\green[1]{{\color{OliveGreen}#1}}

\DeclareMathOperator*{\argmax}{arg\,max}
\DeclareMathOperator*{\argmin}{arg\,min}

\definecolor{green_1}{rgb}{0, 0.8, 0}
\definecolor{green_2}{rgb}{0, 0.6, 0}
\definecolor{green_3}{rgb}{0, 0.4, 0}
\definecolor{red_1}{rgb}{1, 0, 0}
\definecolor{orange_1}{rgb}{1, 0.7, 0}
\definecolor{grey_1}{rgb}{0.8, 0.8, 0.8}
\definecolor{black}{rgb}{0, 0, 0}
\definecolor{lightblue}{rgb}{.7, .7, 1}

\newcommand\legendbox[1]{\textcolor{#1}{\rule[0pt]{6pt}{6pt}}}
\newcommand\legendline[1]{\textcolor{#1}{\rule[1pt]{10pt}{2pt}}}

\algrenewcommand\algorithmicindent{1.0em}%

\journal{Journal of Sound and Vibration}

\begin{document}

\begin{frontmatter}

\title{Mode Selection for Component Mode Synthesis with Guaranteed Assembly Accuracy}

\author[aff1]{Lars A.L. Janssen \corref{corresponding}}
\ead{L.A.L.Janssen@tue.nl}

\author[aff1]{Rob H.B. Fey }
\ead{R.H.B.Fey@tue.nl}

\author[aff2]{Bart Besselink }
\ead{B.Besselink@rug.nl}

\author[aff1]{Nathan van de Wouw}
\ead{N.v.d.Wouw@tue.nl}

\cortext[corresponding]{Corresponding author}
\address[aff1]{Dynamics \& Control, Department of Mechanical Engineering,  Eindhoven University of Technology}
\address[aff2]{Bernoulli Institute for Mathematics, Computer Science and Artificial Intelligence, University of Groningen}

\begin{abstract}
In this work, a modular approach is introduced to select the most important eigenmodes for each component of a composed structural dynamics system to obtain the required accuracy of the reduced-order assembly model.
To enable the use of models of complex (structural) dynamical systems in engineering practice, e.g., in a design, optimization and/or control context, the complexity of the models needs to be reduced.
When the model consist of an assembly of multiple interconnected structural components, component mode synthesis is often the preferred model reduction method.
The standard approach to component mode synthesis for such system is to select the eigenmodes of a component that are most important to accurately model the dynamic behavior of this component in a certain frequency range of interest.
However, often, a more relevant goal is to obtain, in this frequency range, an accurate model of the assembly.
In the proposed approach, accuracy requirements on the level of the assembly are translated to accuracy requirements on component level, by employing techniques from the field of systems and control.
With these component-level requirements, the eigenmodes that are most important to accurately model the dynamic behavior of the assembly can be selected in a modular fashion.
We demonstrate with two structural dynamics benchmark systems that this method based on assembly accuracy allows for a computationally efficient selection of eigenmodes that 1) guarantees satisfaction of the assembly accuracy requirements and 2) results in most cases in reduced-order models of significantly lower order with respect to the industrial standard approach in which component eigenmodes are selected using a frequency criterion.
\end{abstract}

\end{frontmatter}

\section{INTRODUCTION}
The application of (structural) dynamic models includes the analysis, prediction, optimization and/or control of dynamical systems.
However, in all of these applications, models are required that 1) accurately represent the system dynamics and 2) can be used with limited computational effort.
This generally results in a trade-off between the accuracy and complexity of a model.
Model order reduction (MOR) techniques deal with this trade-off by reducing the complexity of an accurate high-dimensional model to obtain a reduced-order model (ROM) that still describes the dynamics of the dynamical system sufficiently accurately. 
This research area spans several interdisciplinary fields, such as structural dynamics, numerical mathematics, and systems and control \cite{besselink2013}.
Typically, MOR methods from these fields have in common that they rely on projection-based methods, such as modal truncation (which is also the reduction mechanism in component mode synthesis methods~\cite{craig1985review}), balancing methods~\citep{moore1981}, and Krylov methods~\citep{grimme1997,freund2003model}. 
These methods work by projecting the high-dimensional model onto a low-order subspace. 

In practice, many models of these complex systems are built by interconnecting multiple component models.
In such systems, the components are often developed by independent teams and, therefore, the modelling of high-dimensional (Finite Element) models of these components is typically also done on component level.
To use the component models for application on a component level, a component ROM can simply be computed using any existing MOR method suitable for the model and application.
However, if the models are to be applied on assembly level, the need arises to take into account the interconnection structure of the interacting components.
This may be done by collecting the high-dimensional component models and constructing a single high-dimensional assembly model \cite{reis2008survey}.
Then, this high-dimensional assembly model is reduced to obtain an assembly ROM that is both of sufficient accuracy and complexity for the desired application.
For this approach, several structure-preserving MOR methods are available that preserve the interconnection structure and component hierarchy of the systems \cite{sandberg2009model,klerk2008}.
However, for many systems, due to their complexity, collecting and coupling all the high-dimensional component models and reducing the resulting assembly model in general is a computationally (very) expensive or even infeasible problem \cite{vaz1990}.

The alternative to this approach is to first reduce the component models independently before connecting these component ROMs in an assembly ROM.
This is computationally significantly cheaper, as by dividing the problem into multiple smaller problems, the reduction of one very high-dimensional model is avoided.
Furthermore, such a modular approach allows for the choice of a different suitable MOR method for each component.
However, when MOR is applied on a component level, generally, an error is introduced in the component ROM with respect to the high-dimensional component model.
As a result, when these component ROMs are assembled to construct an assembly ROM, these errors can potentially propagate and result in an assembly model that is not sufficiently accurate.
Therefore, there is a need for MOR methods that take into account the required assembly accuracy when reducing the complexity of component models.

In the field of structural dynamics, component mode synthesis (CMS) is a projection-based approach that typically involves selecting the most important vibration modes from the component Finite Element (FE) models (often based on a frequency criterion) and then combining them in a way that accurately represents the dynamic behavior of the original assembly.
This approach has been studied since the early 1960s \cite{hurty1960vibrations,hurty1965dynamic} and it remains an area of active research to this day \cite{allen2019substructuring}.
Examples of frequently used CMS methods are the Craig-Bampton~\cite{craig1968coupling}, Hintz-Herting~\cite{hintz1975analytical,herting1985general}, Rubin~\cite{rubin1975improved} and Craig-Chang~\cite{craig1977use} CMS methods, which are also implemented in commercial FE packages. However, guarantees about the accuracy of these assembly ROM are not given.
Reviews of CMS methods are given in \cite{hurty1965dynamic,craig1985review,craig2000coupling,klerk2008,geradin2014mechanical,sun2023review}.
%All of these methods aim to achieve component ROMs that, when interconnected, lead to an accurate assembly ROM.

In engineering practice, accuracy requirements are defined as a frequency range of interest in which the assembly ROM needs to describe the dynamic behaviour of the system with a specified level of accuracy.
However, guaranteeing that such requirements are satisfied when the components are reduced individually is difficult, since predicting the propagation of component ROM errors to the assembly ROM is often not possible since there are generally no a priori error bounds available for CMS methods \cite{seshu1997substructuring}.
Therefore, to make sure that the assembly ROM is sufficiently accurate, the industry standard is to select, for all components,  all vibration modes up to a predefined cut-off frequency.
This cut-off frequency is typically chosen to be around two or three times the maximum frequency of interest of the assembly dynamics \cite{voormeeren2012dynamic,kim2018component,seshu1997substructuring,meirovitch1980computational}. 
% this approach introduces some potential problems.
However, this approach does not take into account the relative importance of component eigenmodes for the assembly model. 
Therefore, often, component modes that are not important for the assembly dynamics are unnecessarily preserved in the ROM. 
In contrast, important (high-frequency) component eigenmodes are potentially not selected even though they are vital to accurately describe the dynamics of the assembly in the frequency range of interest.
%Secondly, because an equal cut-off frequency is typically selected for all components, the relative importance of components is not taken into account, even though some components are potentially much more important to the assembly dynamics than others.

This problem can be tackled with mode selection methods.
In general, with mode selection methods in MOR of structural dynamics systems, the aim is to obtain a selection of eigenmodes for each of the individual components that results in an assembly ROM that satisfies certain accuracy requirements with as few eigenmodes as possible.
In theory, one could apply a brute force method to compute an assembly ROM and compare a posteriori each combination of selected component eigenmodes.
Then, one could pick the selection of modes that fulfils the accuracy requirements and requires the least amount of total eigenmodes.
However, for virtually any practical example, this becomes computationally infeasible.
Therefore, component eigenmode selection methods have been developed that do not require a brute force approach and have shown to significantly outperform the standard of selecting all eigenmodes up to a certain cut-off frequency \cite{kim2016evaluating,sun2023review}.
%There are several mode selection methods available that select the most important modes of components \cite{kim2016evaluating,sun2023review}.
Examples of these methods are the EIM-method \cite{kammer1996selection}, the OMR-method \cite{givoli2004OMR}, the CMS$_{\chi}$-method \cite{liao2007chi}, and sensitivity-based methods \cite{kessels2022sensitivity}.
All of these methods have in common that they aim to select the modes on a component level in such a way that the assembly model is as accurate as possible.
%They all show that they can improve this accuracy with respect to the industrial standard of simply selecting all component modes up to a cut-off frequency.
However, none of these methods can \emph{guarantee} a certain level of accuracy of the assembly ROM a priori.

In the field of systems and control, component-level reduction methods are often referred to as modular, subsystem, or separate bases MOR \cite{vaz1990,schilders2014novel,lutowska2012}.
In this field, similar problems arise when subsystems/components are reduced without taking into account the overall goal of achieving an accurate interconnected system/assembly model.
As a solution, in \cite{janssen2022modular}, a framework was proposed to relate a priori component-level reduction errors to reduction errors on assembly level in the frequency domain. 
This is achieved by formulating this relation in the context of a robust performance analysis problem, which originates from the field of robust control \cite{zhou1998}. 
In \cite{janssen2023modular}, this framework is used to, from the top down, compute accuracy requirements on the input-to-output component dynamics directly based on accuracy specifications on the external input-to-output dynamics of the assembly ROM.
If component ROMs are found that satisfy these ``local'' requirements, the assembly ROM accuracy requirements are guaranteed to also be satisfied.

In \cite{janssen2023translating}, the top-down framework introduced in \cite{janssen2023modular} was applied to the component mode synthesis modelling approach developed for complex structural dynamic systems.
In \cite{janssen2023translating}, component-level cut-off frequencies are adaptively selected for each component such that accuracy requirements on the assembly model are guaranteed to be satisfied.
%The Hintz-Herting CMS-reduction method~\cite{hintz1975analytical} as an example is applied, but it is important to note that any CMS-method could have been applied.
In this work, it is shown that component ROMs can be found that are reduced further in comparison to that with industrial standard while at the same time, it is guaranteed that the assembly ROM accuracy requirements are satisfied.
However, the frequency-dependent component-level requirements obtained from the framework introduced in \cite{janssen2023modular} offer the potential to do much more than to only adaptively choose the cut-off frequency for the reduction of each component model.

The main contribution of the current paper is to employ requirements obtained using the framework introduced in \cite{janssen2023modular} to define an component eigenmode selection approach that allows for even further reduction of the component models while guaranteeing the accuracy of the assembly ROM.
Specifically, we achieve this by introducing the notion of relative mode importance (RMI).
With the RMI, we can compute the effect of adding or removing modes from a component model based on the component-level accuracy requirement.
Using the RMI, we introduce several mode selection approaches that can select the most important eigenmodes of each component, for the satisfaction of the assembly ROM accuracy requirements, in a completely modular approach.

The approach is illustrated on two benchmark systems.
First, we show on a simple model of two interconnected cantilever beams that the industrial standard is often either (much) too conservative or not accurate enough, while the proposed approach automatically takes the relative importance of component modes into account, and, therefore, generally achieves more accurate and less complex models, i.e., ROMs of lower order. 
In addition, we compare the mode selection methods to the optimal selection (determined by an expensive brute force approach) and show that for the cases where computing the optimal selection is feasible, the mode selection methods indeed result in the optimal selection of modes with respect to the component requirements.
We also discuss the advantages and disadvantages of the different mode selection approaches in terms of the number of selected modes and the computational cost for this simple benchmark.
Second, we apply the approach to a three-component structural dynamics system inspired by an industrial wire bonder system and show that the proposed approach allows for the construction of an assembly ROM that 1) is guaranteed to satisfy the specified assembly accuracy requirements, 2) is reduced in a completely modular approach, and 3) is significantly further reduced with respect to the approach given in~\cite{janssen2023translating} and especially with respect to the industrial standard.

The paper is organized as follows. Section~\ref{sec:framework} gives the modelling framework from component discretization and reduction to assembly coupling. 
In Section~\ref{sec:methodology}, we describe how accuracy component requirements can be achieved from accuracy requirements on the assembly-level.
Then, using these component accuracy requirements, the concept of Relative Mode Importance and several mode selection approaches that make use of this concept are introduced in Section~\ref{sec:mode_selection}.
In Section~\ref{sec:examples}, the proposed methods are applied to two different illustrative case studies.
Finally, the conclusions are given in Section~\ref{sec:conclusions}.

\section{Component-based model reduction}
\label{sec:framework}
\begin{figure}
  	\centering
   	\includegraphics[scale=.8, page=1]{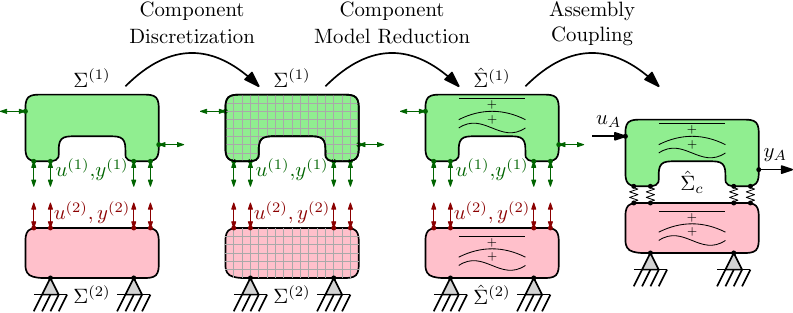}
   	\caption{General approach for model order reduction of a system consisting of multiple components \cite{janssen2023translating}.}
    \label{fig:gen_approach}
\end{figure}
The typical approach used to obtain an assembly ROM that is applicable for design optimization, parameter studies and control applications is illustrated in Figure~\ref{fig:gen_approach}.
In this section, the discretization and reduction of complex component models and their coupling to form an assembly ROM are elaborated.
%This framework follows closely the framework given in \cite{janssen2023translating}.

\subsection{Component Discretization}
In this work, we assume that the assembly consists of $k$ spatially discretized components where each component $j \in \bk:=\{1,\dots,k\}$ can be represented by a high-order, linear, time-invariant, second-order (FE) model of the form
\begin{align}
\label{eq:sigma_j}
\Sigma^{(j)} : M^{(j)}\ddot{q}^{(j)} + C^{(j)} \dot{q}^{(j)} + K^{(j)}q^{(j)} = B^{(j)} u^{(j)}, \quad y^{(j)} = F^{(j)} q^{(j)}. 
\end{align}
Here, the $n_j$ degrees of freedom (DOF) of component $j$ are stored columnwise in $q^{(j)}$, and $M^{(j)}, C^{(j)}$, and $K^{(j)}$ represent the component's mass, damping, and stiffness matrix, respectively, $B^{(j)}$ is the input matrix, and $F^{(j)}$ is the displacement output matrix. 
Furthermore, the frequency response function (FRF) relating the $m_j$ input forces $u^{(j)}$ to the $p_j$ output positions $y^{(j)}$ in the frequency domain is given as
\begin{equation}
\label{eq:H}
H^{(j)}(\im\omega) := F^{(j)} \left(-\omega^2 M^{(j)} + \im\omega C^{(j)} + K^{(j)} \right)^{-1} B^{(j)}.
\end{equation}
\begin{remark}
Obviously, there is an efficient method available to compute the FRF (or dynamic flexibility matrix) of (\ref{eq:sigma_j}) using modal analysis \cite{de2004numerical,geradin2014mechanical,craig2006} where the FRF is constructed by superposition of the contributions of the component's eigenmodes. 
In this work, we will utilize this efficient method to compute how the FRF of a component is affected if a specific eigenmode is excluded or added.
\end{remark}

\subsection{Component Model Reduction}
%As stated before, for complex components, the number of DOF $n_j$ is generally too large for parameter studies, design optimization, and control applications. 
A popular family of methods to compute a reduced-order model $\hat{\Sigma}^{(j)}$ with $r_j \ll n_j$ DOF $\hat{q}^{(j)}$ are component mode synthesis (CMS) methods \cite{craig2000coupling}.
In CMS, the component DOF $q^{(j)}$ are partitioned, i.e., ${q^{(j)}}^\top = \left[\begin{array}{cc} {q^{(j)}_i}^\top & {q^{(j)}_b}^\top\end{array}\right]$, where $q_i$ are internal and $q_b$ are boundary (or interface) DOF. Boundary DOF are (user specified) DOF which are loaded by external loads and/or adjacent components.
The matrices in (\ref{eq:sigma_j}) can be partitioned accordingly, leading to
\begin{align}
M^{(j)} &= \left[\begin{array}{cc}
M^{(j)}_{ii} & M^{(j)}_{ib} \\ M^{(j)}_{bi} & M^{(j)}_{bb} \end{array}\right],& \quad
C^{(j)} &= \left[\begin{array}{cc}
C^{(j)}_{ii} & C^{(j)}_{ib} \\ C^{(j)}_{bi} & C^{(j)}_{bb} \end{array}\right],& \quad
K^{(j)} = \left[\begin{array}{cc}
K^{(j)}_{ii} & K^{(j)}_{ib} \\ K^{(j)}_{bi} & K^{(j)}_{bb} \end{array}\right], \nonumber \\
B^{(j)} &= \left[\begin{array}{c} O \\B^{(j)}_{b}\end{array} \right],& \quad
F^{(j)} &= \left[\begin{array}{cc} O & {F^{(j)}_{b}}\end{array} \right]. 
\end{align} 
%For a more comprehensive overview of these component mode synthesis methods, see \cite{craig2000coupling}.
%In general, each of these methods can be applied to the method that will be proposed.
Note that with the definition of this partitioning, inputs forces and outputs positions are applied to/measured at the boundary DOFs $q^{(j)}_b$.
CMS reduction techniques generally preserve the boundary DOFs $q^{(j)}_b$, as this allows for connection of the reduced-order component models.
For example, in the Hintz-Herting CMS method (HH-MOR)\cite{herting1985general}, the undamped eigenvalue problem $(K^{(j)}-\omega^2_\ell M^{(j)})\phi^{(j)}_\ell = 0$ is solved (using an efficient iterative eigenvalue solver such as for example the Lanczos solver) for $\bar{n}_j$ eigenfrequencies up to a user-defined cut-off frequency, i.e., $0 \leq \omega_\ell \leq  \omega_{\cut}$. 
%In theory, any CMS method that preserves the boundary DOFs $q^{(j)}_b$ can be applied with the methodology that will be proposed in Section~\ref{sec:mode_selection}, e.g., the Craig-Bampton, Rubin, or Hintz-Herting CMS methods. 
%In this work, as an example, the Hintz-Herting CMS method (HH-MOR)\cite{herting1985general} is applied. 
%
%In the HH-MOR method, 
%While the choice of this cut-off frequency in fact defines the reduction of the component DOF in the classic CMS approach, in the current paper, this step only serves as a, generally conservative, preselection (first reduction step) of important eigenmodes. 
%A second reduction step by selecting the $r_j$ most important eigenmodes from the set of preselected eigenmodes will subsequently be performed as will be described in detail later.
The corresponding free-interface eigenmodes $\phi^{(j)}_\ell$ are split into rigid body modes (for which $\omega_\ell = 0$), if any, and elastic free-interface eigenmodes (for which $0 < \omega_\ell \leq \omega_{\cut}$), and, respectively, collected in 
\begin{equation}
{\Phi_r^{(j)}} = \left[\begin{array}{c} {\Phi_{r,i}^{(j)}} \\ {\Phi_{r,b}^{(j)}} \end{array}\right] \text{ and }
{\Phi_e^{(j)}} = \left[\begin{array}{c} {\Phi_{e,i}^{(j)}} \\ {\Phi_{e,b}^{(j)}} \end{array}\right],
\end{equation}
where the matrices $\Phi_r$ and $\Phi_e$ are partitioned according to the internal and boundary DOF.
It is assumed that the components are proportionally or weakly damped, so that use of (real) undamped eigenmodes is valid.
The \emph{static constraint modes} are given by $\Psi^{(j)} := -(K_{ii}^{(j)})^{-1}K^{(j)}_{ib}$, and the \emph{inertia relief modes} (if any, their number is equal to the number of rigid body modes) are given by $\Phi_{ir}^{(j)} := -(K_{ii}^{(j)})^{-1}(M_{ib}^{(j)}+M_{ii}^{(j)}\Psi^{(j)})\Phi_{r,b}^{(j)}$, and the \emph{uncoupled (from $q_b$) elastic eigenmodes} are given by $\Phi_{\epsilon}^{(j)} := \Phi_{e,i}^{(j)} - \Psi^{(j)}\Phi_{e,b}^{(j)}$. 
Then, the component transformation matrix is given by
\begin{align}
\label{eq:T}
T^{(j)} &:= \left[\begin{array}{ccc}
\Phi_{ir}^{(j)} & \Phi_{\epsilon}^{(j)} & \Psi^{(j)} \\
O & O & I
\end{array}\right],
\end{align}
and this matrix gives the relation between the original DOF $q^{(j)}$ and the reduced DOF $\hat{q}^{(j)}$ of the component: $q^{(j)}=T^{(j)} \hat{q}^{(j)}$.
Given a transformation matrix $T^{(j)} \in \R^{n_j \times r_j}$, such as (\ref{eq:T}), the component matrices (\ref{eq:sigma_j}) can be reduced to obtain
\begin{align*}
\hat{M}^{(j)} &= {T^{(j)}}^\top M^{(j)} T^{(j)},& \hat{K}^{(j)} &= {T^{(j)}}^\top K^{(j)} T^{(j)},& \hat{C}^{(j)} &= {T^{(j)}}^\top C^{(j)} T^{(j)}, \\
\hat{B}^{(j)} &= {T^{(j)}}^\top B^{(j)},& \hat{F}^{(j)} &= F^{(j)} T^{(j)},
\end{align*}
and the reduced-order component model for $j \in \bk$ becomes
\begin{equation}
\label{eq:hatsigma_j}
\hat{\Sigma}^{(j)} : \hat{M}^{(j)}\ddot{\hat{q}}^{(j)} + \hat{C}^{(j)} \dot{\hat{q}}^{(j)} + \hat{K}^{(j)}\hat{q}^{(j)} = B^{(j)} \hat{u}^{(j)} , \quad \hat{y}^{(j)} = \hat{F}^{(j)} \hat{q}^{(j)}. 
\end{equation} 
Note that the number of inputs is not reduced, i.e., $\hat{u}^{(j)}$ has the same dimension as $u^{(j)}$; the notation merely indicates that the values of $\hat{u}^{(j)}$ will differ from the values of $u^{(j)}$ after coupling of the reduced component models.
The total number of modes in $\hat{\Sigma}^{(j)}$ is given by the sum of the number of inertia relief modes $r_{j,ir}$, uncoupled elastic modes $r_{j,\epsilon}$, and static constraint modes $r_{j,b}$, i.e., $r_j = r_{j,ir} + r_{j,\epsilon} + r_{j,b}$.
%Note that the number of inertia relief modes equals the rigid body modes and may be zero.
Note that the linear space spanned by the rigid body modes, if any, will be a subspace of the space spanned by the static constraint modes.

Generally, the reduction of the component model $\Sigma^{(j)}$ to $\hat{\Sigma}^{(j)}$ introduces a loss of accuracy.
To quantify this accuracy loss, we introduce the component's error dynamics as
\begin{align}
\label{eq:Ej}
E^{(j)}(\im\omega) := \hat{H}^{(j)}(\im\omega) - H^{(j)}(\im\omega),
\end{align}
for $\omega\in\R$. 
Here, $\hat{H}^{(j)}(\im\omega)$ is the FRF of (\ref{eq:hatsigma_j}) which is defined similarly as in (\ref{eq:H}).
%\begin{remark}
%%The uncoupled elastic eigenmodes $\Phi_{\epsilon}^{(j)}$ can also be constructed from a specific selection of the elastic free-interface eigenmodes $\Phi_{e}^{(j)}$.
%In this work, the aim is to employ a second selection/reduction step starting with the set of eigenmodes with eigenfrequencies in the range $0$ to $\omega_{\cut}$, from which the $r_j$ most important eigenmodes for accurate description of the assembly input-output FRF behavior in a frequency range of interest are preserved in the ROM by selecting specific columns from $\Phi_{\epsilon}^{(j)}$.
%In Section~\ref{sec:mode_selection}, this second reduction step problem is further elaborated in order to realize that the selection of the component eigenmodes is such that the required accuracy of the assembly ROM is realised. 
%To avoid unnecessary complex notation, from now on, we assume that $\Phi^{(j)}_{\epsilon}$ and $\Phi^{(j)}_{e}$ contain the eigenmodes selected after the two reduction steps mentioned above. 
%Note that all other types of modes are kept.
%\end{remark}

\subsection{Assembly Modelling by Flexible Coupling of Components}
\begin{figure}
  	\centering
   	\includegraphics[scale=1, page=2]{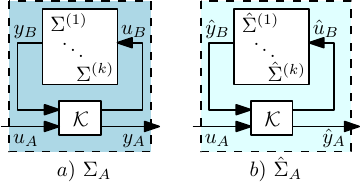}
   	\caption{Assembly coupling of the a) high-order and b) reduced-order component models using the interconnection matrix $\K$.}
    \label{fig:coupling}
\end{figure}

In contrast to the common approach in CMS to \emph{rigidly} couple the components by requiring equilibrium of internal interface forces and compatibility of interface displacements, in this paper, the component couplings will be \emph{flexible} although they still may be (very) stiff to model almost rigid interconnections between components.
To define a general notation of this flexible coupling between components, the component input forces and output positions are combined into $u_B^\top = [{u^{(1)}}^\top,\dots,{u^{(k)}}^\top]$ and  $y_B^\top = [{y^{(1)}}^\top,\dots,{y^{(k)}}^\top]$, respectively, and, similarly for the component ROMs into $\hat{u}_B$ and $\hat{y}_B$.
Then, both the high-order assembly model $\Sigma_A$ and the assembly ROM $\hat{\Sigma}_A$ are formed using the interconnection matrix $\K$, which connects the components $\Sigma^{(j)}$ as in (\ref{eq:sigma_j}) and $\hat{\Sigma}^{(j)}$ as in (\ref{eq:hatsigma_j}) for all $j \in \bk$, respectively. 
This connection is given by
\begin{align}
\label{eq:sigmac}
\left[\begin{array}{c}
u_B \\ y_A 
\end{array}\right] = \K \left[\begin{array}{c}
y_B \\ u_A 
\end{array}\right] \text{ and } \left[\begin{array}{c}
\hat{u}_B \\ \hat{y}_A 
\end{array}\right] = \K \left[\begin{array}{c}
\hat{y}_B \\ u_A 
\end{array}\right] \text{ with } \K = \left[\begin{array}{cc}
\K_{BB} & \K_{BA} \\
\K_{AB} & O
\end{array}\right],
\end{align}
respectively.
Here, $u_A$ are the $m_A$ external input forces, and $y_A$ and $\hat{y}_A$ are the $p_A$ external output positions of $\Sigma_A$ and $\hat{\Sigma}_A$, respectively.
The total number of DOF in $\Sigma_A$ is given by $n_A := \sum_{j=1}^k n_j$ and in $\hat{\Sigma}_A$ by $r_A := \sum_{j=1}^k r_j$.
In Figure~\ref{fig:coupling}, the coupling approach is illustrated for both the a) high-order and b) reduced-order assembly models.
%\begin{remark}
%In this work, it is assumed that components are connected through elastic interfaces.
%However, these interfaces can potentially be made very stiff to model (almost) rigid interconnections between components.
%\end{remark}
We define $H_B := \text{diag}(H^{(1)},\dots,H^{(k)})$ and $\hat{H}_B := \text{diag}(\hat{H}^{(1)},\dots,\hat{H}^{(k)})$.
Then, the FRFs from $u_A$ to $y_A$ and from $u_A$ to $\hat{y}_A$ are given by
\begin{align}
\label{eq:Gc}
H_A(\im\omega) &:= \K_{AB}H_B(\im\omega)\left( I - \K_{BB}H_B(\im\omega)\right)^{-1}\K_{BA}, \text{ and }\\ 
\hat{H}_A(\im\omega) &:= \K_{AB}\hat{H}_B(\im\omega)\left( I - \K_{BB}\hat{H}_B(\im\omega)\right)^{-1}\K_{BA},
\end{align}
respectively. 
The assembly model \emph{error dynamics}, i.e., the difference between the high-order and reduced-order assembly model FRFs, are then given by
\begin{align}
\label{eq:Ec}
E_A(\im\omega) &:= \hat{H_A}(\im\omega)-H_A(\im\omega) \nonumber \\ 
&= \K_{AB}\left(\hat{H}_B(\im\omega)\left( I-\K_{BB}\hat{H}_B(\im\omega)\right)^{-1}-H_B(\im\omega)\left( I-\K_{BB}H_B(\im\omega)\right)^{-1}\right)\K_{BA},
\end{align}
for $\omega \in \R$.

\section{DERIVING COMPONENT ACCURACY REQUIREMENTS FROM ASSEMBLY ACCURACY REQUIREMENTS}
\label{sec:methodology}
\begin{figure}
  	\centering
   	\includegraphics[scale=.7, page=3]{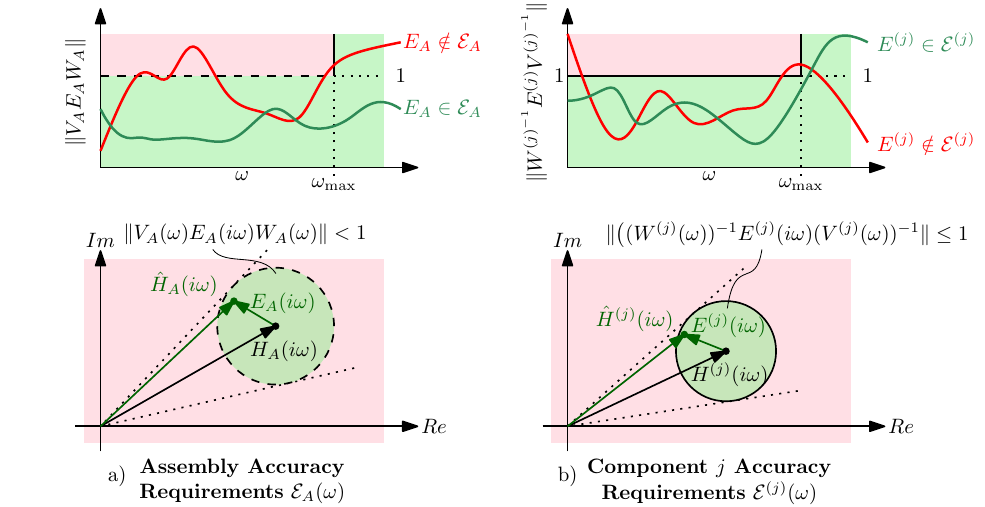}
   	\caption{Visualization of a) the assembly accuracy requirements $\E_A(\omega)$, as given in (\ref{eq:Ec_bound}), and b) the component accuracy requirements $\E^{(j)}(\omega)$, as given in (\ref{eq:Ej_bound}). The scaled errors, i.e., errors scaled by the weighting matrices ($V_A(\omega)$, $W_A(\omega)$, $V^{(j)}(\omega)$, and $W^{(j)}(\omega)$), that remain in the green area satisfy the requirements for all $\omega \in \Omega$. The ROM accuracy requirements are visualised as a function of $\omega$ (top figures) and in the complex plane for any specific $\omega\in\Omega$ (bottom figures). Note that the bottom figure visualizes a SISO system.}
    \label{fig:reqs}
\end{figure}

For most mechanical systems, we are interested in models that are accurate up to a certain frequency.
We define this \emph{maximum frequency of interest} by $\omega_{\max}$ [rad/s] or by $f_{\max} = \omega_{\max} / (2\pi)$ [Hz].
In addition, we define the \emph{frequency range of interest} by $\Omega := \{\omega \ | \ 0 \leq \omega \leq \omega_{\max} \}$.
Note that to reduce the computational load of the approach, frequencies $\omega$ are typically selected on a discrete grid of $\Omega$, defined by $\hat{\Omega}$, for which holds that $\hat{\Omega} \subset \Omega$.
Let us introduce a frequency-dependent assembly accuracy requirement $\E_A(\omega)$, expressed by the set of acceptable assembly ROM errors
\begin{align}
\label{eq:Ec_bound}
\E_A(\omega) := \big\{ E_A(\im\omega) \ \big| \ \|V_A(\omega)E_A(\im\omega)W_A(\omega)\| < 1 \big\}.
\end{align}
Here, $E_A(\im\omega) \in \E_A(\omega)$ implies that the error $E_A(\im\omega)$ introduced by constructing the assembly ROM in (\ref{eq:Ec}) satisfies the assembly ROM accuracy requirements.
In this formulation, the set of acceptable assembly ROM errors $\E_A(\omega)$ is based upon diagonal, frequency-dependent scaling matrices $V_A(\omega) \in \R^{p_A \times {p_A}}_{>0}$ and $W_A(\omega) \in \R^{m_A \times {m_A}}_{>0}$. 
As such, $W_A(\omega)$ and $V_A(\omega)$ in fact define the assembly accuracy requirements.
A visualization of the assembly ROM accuracy requirement $\E_A(\omega)$ is given in Figure~\ref{fig:reqs}a.
For any assembly ROM for which it holds that $E_A(\im\omega)\in\E_A(\omega)$ for all $\omega\in\Omega$, we denote that $E_A\in\E_A$.
In particular, because $\E_A(\omega)$ is frequency-dependent, the accuracy requirement of the assembly ROM (and thus $W_A(\omega)$ and $V_A(\omega)$) can be specifically tailored to its application, i.e., at each $\omega\in\Omega$, $W_A(\omega)$ and $V_A(\omega)$ can be used to impose different requirements on the accuracy of the assembly ROM in terms of the external input-output pairs of $E_A(\im\omega)$.
These requirements can be for example defined by the system engineers, control engineers, or end-users of the model.
For example, the design of $W_A(\omega)$ and $V_A(\omega)$ to impose a requirement on the maximum acceptable assembly ROM \emph{relative error} is given in Section~\ref{sec:examples}.

Given requirements on the assembly ROM accuracy in the form of (\ref{eq:Ec_bound}), we will construct accuracy requirements for the component ROMs (in similar form) given by the sets of acceptable errors for the ROM of each component $j \in \bk$:
\begin{align}
\label{eq:Ej_bound}
\E^{(j)}(\omega) := \big\{ E^{(j)}(\im\omega) \ \big| \ \| (W^{(j)}(\omega))^{-1}E^{(j)}(\im\omega)(V^{(j)}(\omega))^{-1} \| \leq 1 \big\},
\end{align}
for any $\omega \in \Omega$, which will be shown to guarantee that $E_A(\im\omega) \in \E_A(\omega)$, and where $V^{(j)}(\omega)$ and $W^{(j)}(\omega)$ are diagonal, frequency-dependent scaling matrices.
A visualization of the component ROM accuracy requirements $\E^{(j)}(\omega)$ is given in Figure~\ref{fig:reqs}b.
For any assembly ROM for which it holds that $E^{(j)}(\im\omega)\in\E^{(j)}(\omega)$ for all $\omega\in\Omega$, we denote that $E^{(j)}\in\E^{(j)}$.
\begin{remark}
Both the component and the assembly requirements are based on a bound on the (scaled) \emph{magnitude} of the error FRFs.
However, please note that a bound on $\|E^{(j)}(\im\omega)\|$ or $\|E_A(\im\omega)\|$ can still impose a bound on the \emph{phase} difference between the high-order and reduced-order FRFs.
For a single-input-single-output (SISO) system, this is demonstrated in Figure~\ref{fig:reqs}. Here, the imposed bound on the phase is illustrated using the dashed lines in the bottom two figures.
\end{remark}

To find (\ref{eq:Ej_bound}), i.e., to find $W^{(j)}$ and $V^{(j)}$, given (\ref{eq:Ec_bound}), the robust performance based methods developed in \cite{janssen2022modular,janssen2023modular} are invoked.
Namely, based on~\cite[Theorem 1]{janssen2023modular}, we obtain the following result:
\begin{theorem}
\label{the:top_down}
Let $\omega\in\R$ and consider the optimization problem 
\begin{align}
\label{eq:top_down}
\textrm{given} \quad & V_A(\omega), W_A(\omega) \nonumber \\
\textrm{minimize} \quad & \tr\left(V^{-2}(\omega)\right) + \tr\left(W^{-2}(\omega)\right) \nonumber \\
\textrm{subject to} \quad & \left[\begin{array}{cc}
W^{-2}(\omega)D_r^{-1} 	& 	N^H(\im\omega) \nonumber \\
N(\im\omega)				& 	V^{-2}(\omega)D_\ell
\end{array}\right] \succ 0, \nonumber \\
\quad & V(\omega) = \textrm{diag}\left(V^{(1)}(\omega),\dots,V^{(k)}(\omega),V_A(\omega)\right) \in \mathbf{V}, \\
\quad & W(\omega) = \textrm{diag}\left(W^{(1)}(\omega),\dots,W^{(k)}(\omega),W_A(\omega)\right) \in \mathbf{W}, \\
\quad & (D_\ell, D_r) \in \mathbf{D},
\end{align}
where 
\begin{align}
\label{eq:N}
N(\im\omega)&:=\left[\begin{array}{cc}
\K_{BB}(I-H_B(\im\omega)\K_{BB})^{-1} & (I-\K_{BB}H_B(\im\omega))^{-1} \K_{BA}\\ 
\K_{AB}(I-H_B(\im\omega)\K_{BB})^{-1}&O
\end{array}\right], \\
\mathbf{V} &:= \Big\{\textrm{diag}(v) \ \Big| \ v \in \R^{m_1+\cdots+m_k+p_A}_{>0} \Big\}, \quad
\mathbf{W} := \Big\{\textrm{diag}(w) \ \Big| \ w \in \R^{p_1+\cdots+p_k+m_A}_{>0} \Big\}, \\
\mathbf{D} &:= \Big\{ (D_\ell, D_r) \ \Big| \ d_1,\dots,d_k, d_A \in \R_{>0}, \ D_\ell = \textrm{diag}\left( d_1 I_{p_1},\dots,d_{k} I_{p_k}, d_AI_{m_A} \right), \\ \nonumber
& \qquad \qquad D_r = \textrm{diag}\left( d_1 I_{m_1},\dots,d_{k} I_{m_k}, d_AI_{p_A} \right)\Big\}.
\end{align}
Consider any feasible solution $V(\omega)$, $W(\omega)$. % and partition this as
%\begin{align}
%V(\omega) = \textrm{diag}(V^{(1)}(\omega),\dots,V^{(k)}(\omega),V_A(\omega)), \quad
%W(\omega) = \textrm{diag}(W^{(1)}(\omega),\dots,W^{(k)}(\omega),W_A(\omega)).
%\end{align}
Then, the following implication holds: if $E^{(j)}(\im\omega)\in \E^{(j)}(\omega)$ for all $j\in \bk$, then $E_A(\im\omega) \in \E_A(\omega)$.
%One such set of requirements is obtained by solving the following optimization problem for $\omega \in \Omega$:
\begin{proof}
%Note that $\tr(V^{-2}(\omega)) + \tr(W^{-2}(\omega)) = \sum_{j=1}^k\tr\left(\left(V^{(j)}(\omega)\right)^{-2}\right) +\tr\left(\left(W^{(j)}(\omega)\right)^{-2}\right)$.
In \cite[Theorem 1]{janssen2023modular} it is proven that any solution to the matrix inequality in (\ref{eq:top_down}) guarantees that $E^{(j)}(\im\omega) \in \E^{(j)}(\omega)$ for all $j\in\bk$ implies $E_A(\im\omega) \in \E_A(\omega)$. 
Furthermore, by minimizing $\tr(V^{-2}(\omega)) + \tr(W^{-2}(\omega))$, automatically, a set of such component ROM accuracy requirements $\E^{(j)}(\omega)$ will be obtained, i.e. matrices $V^{(j)}$ and $W^{(j)}$ will be obtained for all $j \in \bk$.
\end{proof}
\end{theorem}
Given assembly ROM accuracy requirements $\E_A(\omega)$ as in (\ref{eq:Ec_bound}), using Theorem~\ref{eq:top_down}, a set of component ROM accuracy requirements $\E^{(j)}(\omega)$ as in (\ref{eq:Ej_bound}) can be found, for which it holds that $E_A(\im\omega) \in \E_A(\omega)$ if $E^{(j)}(\im\omega) \in \E^{(j)}(\omega)$ for all $j\in\bk$.
Additionally, in Theorem~\ref{the:top_down}, by minimizing $\tr\left(V^{-2}(\omega)\right) + \tr\left(W^{-2}(\omega)\right)$, given a fixed $V_A(\omega)$ and $W_A(\omega)$, the weights $V^{(j)}$ and $W^{(j)}$ for all $j\in\bk$ used for $\E^{(j)}(\omega)$ are maximized.
By doing this, automatically, component ROM accuracy requirements $\E^{(j)}(\omega)$ for all $j\in\bk$ are obtained where components that have a small influence on the external input-output FRF accuracy of the assembly are given more room for errors while components that are more critical for this accuracy are required to be more accurate.
\begin{remark}
In (\ref{eq:top_down}), apart from $V^{(j)}$ and $W^{(j)}$, the additional decision variables $D_\ell$ and $D_r$ are scaling matrices originating from computing an upper bound on the structured singular value, which is used in robust performance analysis to obtain performance bounds and guarantees on the behaviour of a system in the presence of uncertainties \cite{packard1993,zhou1998}. 
This analysis forms the basis for the results obtained in \cite{janssen2022modular} and \cite{janssen2023modular}, which in turn are used in Theorem~\ref{the:top_down}.
In \cite{janssen2023modular}, it is shown how the optimization problem can be efficiently solved through an iterative approach using two linear matrix inequalities obtained by alternating between fixing $V$ and $W$ and fixing $D_\ell$ and $D_r$ until the convergence criteria are met.
\end{remark}
%In Section~\ref{sec:examples}, we show by means of two case studies that this approach for determining component accuracy requirements based on assembly accuracy requirements allows for a further reduction of the assembly model as a whole.

With Theorem~\ref{the:top_down}, an assembly ROM  can be efficiently computed that satisfies the accuracy requirements $\E_A$ for the external input-output FRFs of the assembly even though the reduction is done on component level in an entirely modular fashion.
Namely, given a user-specified $\E_A(\omega)$ for all $\omega \in \Omega$, the optimization problem (\ref{eq:top_down}) can be solved to find $\E^{(j)}$ as in (\ref{eq:Ej_bound}) for all $j \in \bk$. 
Given these accuracy requirements for all components $j \in \bk$, component ROMs $\hat{\Sigma}^{(j)}$ can be independently computed using any suitable CMS method (or reduction method in general) by selecting the necessary eigenmodes $\Phi^{(j)}$ required to meet its component accuracy requirements $\E^{(j)}$. 
To find these necessary component eigenmodes, in Section~\ref{sec:mode_selection}, different mode selection methods are introduced.

\section{MODE SELECTION WITH ASSEMBLY ACCURACY GUARANTEES}
\label{sec:mode_selection}
In the previous section, it is shown how, in a frequency range of interest, assembly ROM accuracy requirements $\E_A$ can be directly translated to accuracy requirements for each component ROM $\E^{(j)}$.
In this section, we will show how the obtained frequency-dependent accuracy requirements $\E^{(j)}$ can be exploited to modularly select eigenmodes of the \emph{components} that are most important to obtain the required accuracy of the \emph{assembly} ROM.
The satisfaction of accuracy requirements only needs to be checked on a component level and can therefore be done with relatively low computational cost.
As a result, mode selection approaches that would otherwise be computationally infeasible now become computationally tractable.

As mentioned in Section~\ref{sec:framework}, we will use the HH-MOR method to first obtain all eigenmodes up to a certain frequency of interest $\omega_{\cut}$.
Namely, all $\bar{n}_{j,\epsilon}$ obtained elastic free-interface modes are collected in the set\footnote{Note that $\Phi_\epsilon^{(j)}$ is defined as a matrix in Section~\ref{sec:framework}, whereas in this section, $\mathbf{\Phi}_\epsilon^{(j)}$ is defined as a set whose elements are the columns of the matrix $\Phi_\epsilon^{(j)}$.}
\begin{align}
	\mathbf{\Phi}_\epsilon^{(j)} = \{\phi_{\epsilon,1}^{(j)},\ldots,\phi_{\epsilon,\bar{n}_{j,\epsilon}}^{(j)}\}. 
\end{align}
Then, a component eigenmode selection can be defined as a subset $\mathbf{\Phi}^{(j)} \subset \mathbf{\Phi}_\epsilon^{(j)}$ of the obtained modes.
When $T^{(j)}$ in (\ref{eq:T}) is constructed using such a subset $\mathbf{\Phi}^{(j)}$ of $\mathbf{\Phi}_\epsilon^{(j)}$, the obtained component ROM is denoted by $\hat{\Sigma}_{\mathbf{\Phi}}^{(j)}$.
Finally, the FRF of $\hat{\Sigma}_{\mathbf{\Phi}}^{(j)}$ is denoted by $\hat{H}_{\mathbf{\Phi}}^{(j)}(\im\omega)$ and, accordingly, 
\begin{align}
E^{(j)}_{\mathbf{\Phi}}(\im\omega) = \hat{H}_{\mathbf{\Phi}}^{(j)}(\im\omega) - H^{(j)}(\im\omega).
\end{align}
The aim of this work is to obtain the smallest selection of component eigenmodes, i.e., a subset $\mathbf{\Phi}^{(j)} \subset \mathbf{\Phi}_\epsilon^{(j)}$, satisfying $E^{(j)}_{\mathbf{\Phi}}(\im\omega) \in \E^{(j)}(\omega)$ for all $\omega \in \Omega$. Following Theorem~\ref{the:top_down}, if such a selection is found for all components $j\in\bk$, then the assembly accuracy requirement is also satisfied, i.e., $E_A(\im\omega) \in \E_A(\omega)$ for all $\omega \in \Omega$.
%In this work, the aim is to employ a second selection/reduction step starting with the set of eigenmodes with eigenfrequencies in the range $0$ to $\omega_{\cut}$, from which the $r_j$ most important eigenmodes for accurate description of the assembly input-output FRF behavior in a frequency range of interest are preserved in the ROM by selecting specific columns from $\mathbf{\Phi}_{\epsilon}^{(j)}$.
%In Section~\ref{sec:mode_selection}, this second reduction step problem is further elaborated in order to realize that the selection of the component eigenmodes is such that the required accuracy of the assembly ROM is realised. 
%To avoid unnecessary complex notation, from now on, we assume that $\mathbf{\Phi}^{(j)}_{\epsilon}$ and $\mathbf{\Phi}^{(j)}_{e}$ contain the eigenmodes selected after the two reduction steps mentioned above. 
%Note that all other types of modes are kept. 
\begin{remark}
There are many possible approaches to select component eigenmodes using the accuracy requirements $\E^{(j)}$ on component level.
By definition, as already mentioned in the introduction, the optimal eigenmode selection $\mathbf{\Phi}^{(j)}$ such that $E^{(j)}_{\mathbf{\Phi}}(\im\omega) \in \E^{(j)}(\omega)$ for all $\omega \in \Omega$ can be found by applying a \emph{brute force} approach to check all combinations of eigenmodes $\mathbf{\Phi}^{(j)} \subset \mathbf{\Phi}_\epsilon^{(j)}$ and select the smallest set of modes which satisfies the accuracy requirements.
However, this approach quickly becomes infeasible as the number of possible options increases exponentially when the number of possible modes increases.
For example, to find the best selection of 10 modes out of 100 possible modes, around $10^{13}$ combinations need to be checked.
Therefore, the need arises for mode selection algorithms that can (sub-optimally) select the most important component eigenmodes much more efficiently.
\end{remark}
To efficiently find an eigenmode selection $\mathbf{\Phi}^{(j)}$ such that $E^{(j)}_{\mathbf{\Phi}}(\im\omega) \in \E^{(j)}(\omega)$ for all $\omega \in \Omega$, we apply algorithms in which we need to add or remove modes from a current selection of modes $\mathbf{\Phi}^{(j)}$.
Therefore, for a given set $\mathbf{\Phi}^{(j)}$, adding or removing a mode $\phi^{(j)}$ from or to a set $\mathbf{\Phi}^{(j)}$ can be written (in pseudo-code) as
\begin{align}
	\mathbf{\Phi}^{(j)} \leftarrow \mathbf{\Phi}^{(j)}\cup\{\phi^{(j)}\} \quad\text{or}\quad \mathbf{\Phi}^{(j)} \leftarrow \mathbf{\Phi}^{(j)}\setminus\{\phi^{(j)}\},
\end{align}
respectively.
An efficient mode selection approach, which we call the frequency-ordered approach, is the component eigenmode selection approach used in \cite{janssen2023translating}.
%Here, for each component an optimal cut-off frequency is determined, i.e., all component eigenmodes up to the optimal cut-off frequency for each specific component are selected.

\subsection{Frequency-ordered Component Eigenmode Selection}
In the frequency-ordered component eigenmode selection approach, as introduced in \cite{janssen2023translating}, the elastic free-interface eigenmodes are added in order of increasing eigenfrequency until the component requirement is satisfied.
For this approach, we denote the eigenfrequency associated to $\phi^{(j)}$ as $\omega(\phi^{(j)})$. 
Then, the frequency-ordered mode selection approach is explicitly described in Algorithm~\ref{alg:Freqordered}.
In addition, in Figure~\ref{fig:freqord}, the approach is illustrated on a simple fictitious example where the subscript $i$ in $\hat{\Sigma}_i^{(j)}$ and $E_i^{(j)}$ denotes the $i$-th iteration.

\begin{tabular}{cc}
\hspace{-.8cm}
\begin{minipage}{.6\textwidth}
\begin{algorithm}[H]
\caption{Frequency-ordered selection}\label{alg:Freqordered}
\textbf{Input: } $\Sigma^{(j)}, \ \mathbf{\Phi}_\epsilon^{(j)}, \ \E^{(j)}(\omega)$ for $\omega \in \Omega$
\begin{algorithmic}[1]
	\State $\mathbf{\Phi}^{(j)} = \emptyset$
	\Repeat
	\State $\phi^{(j)}\leftarrow\argmin\limits_{\phi^{(j)}\in\mathbf{\Phi}_e^{(j)}\setminus\mathbf{\Phi}^{(j)}} \omega(\phi^{(j)})$
	\State $\mathbf{\Phi}^{(j)}\leftarrow\mathbf{\Phi}\cup\{\phi^{(j)}\}$
	\Until{$\hat{E}^{(j)}_{\mathbf{\Phi}}(\im\omega)\in \mathcal{E}^{(j)}(\omega)$ for all $\omega\in\Omega$}
\end{algorithmic}
\textbf{Output:} $\mathbf{\Phi}^{(j)}$
\end{algorithm}
\end{minipage} 
\hspace{-.4cm}&\hspace{-.4cm}
\begin{minipage}{.4\textwidth}
   	\includegraphics[scale=.6, page=2]{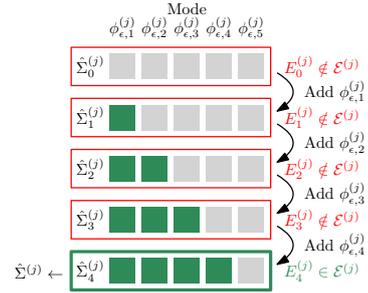}
   	\captionof{figure}{Freq.-ordered mode selection algorithm.}
    \label{fig:freqord}
\end{minipage}\hspace{-.8cm}
\end{tabular}

%In the current paper, we introduce four component eigenmode selection methods, which are more efficient than the approach used in \cite{janssen2023translating}.
%These methods rely on the introduction of the \emph{relative mode importance} (RMI) with respect to the component accuracy requirement $\E^{(j)}$.
%Similar to the frequency-ordered method, these methods aim to find sets of component eigenmodes in which the accuracy requirements $E^{(j)}(\omega) \in \E^{(j)}(\omega)$ for all $\omega \in \Omega$ are satisfied.
%However, in the current paper, component eigenmodes that are relatively more important for satisfaction of component accuracy requirements $\E^{(j)}$ are prioritized.
%In this section, these four component eigenmode selection methods, which are called RMI-A (two variants) and RMI-R (also two variants) are described and illustrated.

\subsection{Component Eigenmode Selection using Relative Mode Importance}
The frequency-ordered method from \cite{janssen2023translating} is a natural choice of mode selection method, since it allows for choosing the lowest cut-off frequency for each component $j$ for which $E^{(j)}(\im\omega) \in \E^{(j)}(\omega)$ is satisfied for all $\omega \in \Omega$.
However, by translating assembly accuracy requirements $\E_A(\omega)$ to component accuracy requirements $\E^{(j)}(\omega)$, explicit frequency-dependent requirements are given on the level of a component.
These requirements can be used to select component eigenmodes that contribute the most to satisfying these requirements, which do not have to be the component eigenmodes corresponding to the lowest eigenfrequencies.
Therefore, we define the notion of relative mode importance (RMI).
\begin{definition}
\label{def:RMI}
The effect on the component ROM accuracy requirements $\E^{(j)}$ of either \emph{adding} or \emph{removing} eigenmode $\phi^{(j)}$ from given selection of modes $\mathbf{\Phi}^{(j)}$ can defined by its relative mode importance (RMI) as follows:
\begin{align}
\label{eq:RMI-A}
	\operatorname{RMI-A}(\mathbf{\Phi}^{(j)},\phi^{(j)}) = \max_{\omega\in\Omega} \, \Bigl|
	&\bigl\|(W^{(j)}(\omega))^{-1}E^{(j)}_{\mathbf{\Phi}} (\im\omega)(V^{(j)}(\omega))^{-1}\bigr\| \\ &- \bigl\|(W^{(j)}(\omega))^{-1}E^{(j)}_{\mathbf{\Phi}\cup\{\phi\}}(\im\omega)(V^{(j)}(\omega))^{-1}\bigr\|
	\Bigr|, \text{ and} \\
\label{eq:RMI-R}
	\operatorname{RMI-R}(\mathbf{\Phi}^{(j)},\phi^{(j)}) = \max_{\omega\in\Omega} \, \Bigl|
	&\bigl\|(W^{(j)}(\omega))^{-1}E^{(j)}_{\mathbf{\Phi}}(\im\omega)(V^{(j)}(\omega))^{-1}\bigr\| \\ &- \bigl\|(W^{(j)}(\omega))^{-1}E^{(j)}_{\mathbf{\Phi}\setminus\{\phi\}}(\im\omega)(V^{(j)}(\omega))^{-1}\bigr\|
	\Bigr|,
\end{align}
respectively. 
\end{definition}
\begin{remark}
There are many possible metrics to determine the importance of an eigenmode to achieve a certain model accuracy.
In this work, we are interested in selecting the component eigenmodes that are most important to satisfy the component accuracy requirement $\E^{(j)}(\omega)$ as introduced in (\ref{eq:Ej_bound}), for all $\omega \in \Omega$, which implies
\begin{align}
\max_{\omega\in\Omega} \, \| (W^{(j)}(\omega))^{-1}E^{(j)}(\im\omega)(V^{(j)}(\omega))^{-1} \| \leq 1.
\end{align}
Therefore, in Definition~\ref{def:RMI}, the RMI is defined such that the more influence a component eigenmode has on the component accuracy requirements $\E^{(j)}(\omega)$ for all $\omega \in \Omega$, the higher the value of $\text{RMI-A}$ or $\text{RMI-R}$.
Specifically, the value of $\text{RMI-A}(\mathbf{\Phi}^{(j)},\phi^{(j)})$ or $\text{RMI-R}(\mathbf{\Phi}^{(j)},\phi^{(j)})$ indicates how much the addition to removal of mode $\phi^{(j)}$ influences the dynamics of a component ROM $\hat{\Sigma}_{\mathbf{\Phi}}^{(j)}$ with respect to the component accuracy requirement $\E^{(j)}$.
\end{remark}
With a computable RMI as in (\ref{eq:RMI-A}), (\ref{eq:RMI-R}), four additional algorithms are introduced that utilize the RMI to prioritize modes with a higher RMI when constructing a component ROM; two methods relying on RMI-A and two methods relying on the RMI-R.

In both RMI-A methods, instead of selecting the eigenmodes corresponding to the lowest eigenfrequency, the eigenmodes corresponding to the highest RMI-A are selected first.
The RMI-A values can be computed \emph{a priori}, i.e, the RMI-A values are computed only once for all modes in the preselected set $\mathbf{\Phi}_\epsilon^{(j)}$ elastic free-interface eigenmodes which are individually added to a component ROM without any elastic eigenmodes, i.e., $\mathbf{\Phi}^{(j)} = \emptyset$. 
This algorithm is given in Algorithm~\ref{alg:RMIadditive1} and illustrated on a fictitious example in Figure~\ref{fig:add_aprio}.
In addition, the RMI-A values can also be computed using an \emph{incremental} approach, i.e., each time a new eigenmode is added to the component ROM, the RMI-A values are computed with respect to the current selection of modes $\mathbf{\Phi}^{(j)}$. 
This algorithm is given in Algorithm~\ref{alg:RMIadditive2} and illustrated on a fictitious example in Figure~\ref{fig:add_greedy}.

\begin{minipage}{.5\textwidth}
\centering
\begin{algorithm}[H]
\caption{RMI-A A priori selection}\label{alg:RMIadditive1}
\textbf{Input: } $\Sigma^{(j)}, \ \mathbf{\Phi}_\epsilon^{(j)}, \ \E^{(j)}(\omega)$ for $\omega \in \Omega$
\begin{algorithmic}[1]
\State $\mathbf{\Phi}^{(j)} = \emptyset$
\Repeat
\State $\phi^{(j)}\leftarrow\argmax\limits_{\phi\in\mathbf{\Phi}_\epsilon^{(j)}\setminus\mathbf{\Phi}^{(j)}} \operatorname{RMI-A}(\emptyset,\phi^{(j)})$
\State $\mathbf{\Phi}^{(j)}\leftarrow\mathbf{\Phi}^{(j)}\cup\{\phi^{(j)}\}$
\Until{\green{$\hat{E}^{(j)}_{\mathbf{\Phi}}(\im\omega)\in \mathcal{E}^{(j)}(\omega)$} for all $\omega\in\Omega$}
\end{algorithmic}
\textbf{Output:} $\mathbf{\Phi}^{(j)}$
\end{algorithm}
%\vspace{.1cm}
\end{minipage} 
\begin{minipage}{.5\textwidth}
\centering
\begin{algorithm}[H]
\caption{RMI-A Incremental selection}\label{alg:RMIadditive2}
\textbf{Input: } $\Sigma^{(j)}, \ \mathbf{\Phi}_\epsilon^{(j)}, \ \E^{(j)}(\omega)$ for $\omega \in \Omega$
\begin{algorithmic}[1]
\State $\mathbf{\Phi}^{(j)} = \emptyset$
\Repeat
\State $\phi^{(j)}\leftarrow\argmax\limits_{\phi^{(j)}\in\mathbf{\Phi}_\epsilon^{(j)}\setminus\mathbf{\Phi}^{(j)}} \operatorname{RMI-A}(\mathbf{\Phi}^{(j)},\phi^{(j)})$
\State $\mathbf{\Phi}^{(j)}\leftarrow\mathbf{\Phi}^{(j)}\cup\{\phi^{(j)}\}$
\Until{\green{$\hat{E}^{(j)}_{\mathbf{\Phi}}(\im\omega)\in \mathcal{E}^{(j)}(\omega)$} for all $\omega\in\Omega$}
\end{algorithmic}
\textbf{Output:} $\mathbf{\Phi}^{(j)}$
\end{algorithm}
%\vspace{.1cm}
\end{minipage}\\
\hspace{.2cm}
\begin{minipage}{.5\textwidth}
	\centering
   	\includegraphics[scale=.8, page=3]{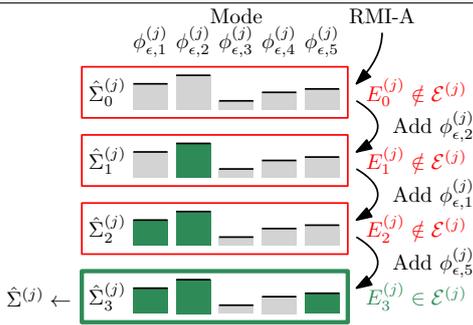}
   	\captionof{figure}{RMI-A A priori approach: Compute RMI-A (indicated by the height of the bars) of all eigenmodes initially, add eigenmodes until \green{$E^{(j)}\in\E^{(j)}$}.}
    \label{fig:add_aprio}
\end{minipage} 
\hspace{.5cm}
\begin{minipage}{.5\textwidth}
	\centering
   	\includegraphics[scale=.8, page=4]{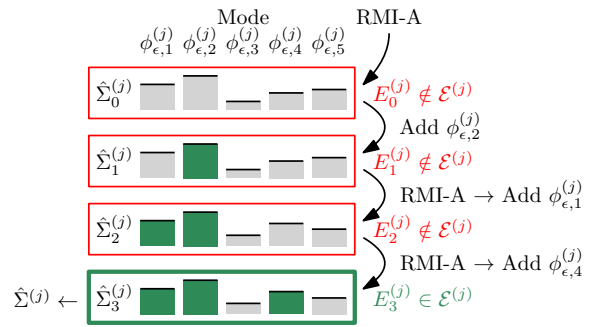}
   	\captionof{figure}{RMI-A Incremental approach: Compute RMI-A of all eigenmodes at every iteration, add eigenmodes until \green{$E^{(j)}\in\E^{(j)}$}.}
    \label{fig:add_greedy}
\end{minipage}

The RMI-R methods, on the other hand, rely on \emph{removing} modes from the component ROM until the component accuracy requirement is no longer satisfied.
In this approach, the eigenmodes corresponding to the lowest RMI-R are removed first.
The RMI-R values can be computed \emph{a priori}, i.e, the RMI-R values are computed only once for all $\mathbf{\Phi}_\epsilon^{(j)}$ elastic free-interface eigenmodes which are individually removed from a component ROM with all elastic eigenmodes, i.e., all $\mathbf{\Phi}_\epsilon^{(j)}$ eigenmodes up to the cut-off frequency which was used in the preselection.
This algorithm is given in Algorithm~\ref{alg:RMIsub1} and illustrated on a fictitious example in Figure~\ref{fig:sub_aprio}.
The RMI-R values can also be computed with an \emph{incremental} approach, i.e., each time a new eigenmode is removed from the component ROM, the RMI-R values are computed with respect to the current selection of modes $\mathbf{\Phi}^{(j)}$. 
This algorithm is given in Algorithm~\ref{alg:RMIsub2} and illustrated on a fictitious example in Figure~\ref{fig:sub_greedy}.

\begin{minipage}{.5\textwidth}
\centering
\begin{algorithm}[H]
\caption{RMI-R A priori selection}\label{alg:RMIsub1}
\textbf{Input: } $\Sigma^{(j)}, \ \mathbf{\Phi}_\epsilon^{(j)}, \ \E^{(j)}(\omega)$ for $\omega \in \Omega$
\begin{algorithmic}[1]
		\State $\mathbf{\Phi}^{(j)} = \mathbf{\Phi}_\epsilon^{(j)}$
		\Repeat
		\State $\phi^{(j)}\leftarrow\argmin\limits_{\phi^{(j)}\in\mathbf{\Phi}^{(j)}} \operatorname{RMI-R}(\mathbf{\Phi}_\epsilon^{(j)},\phi^{(j)})$
		\State $\mathbf{\Phi}^{(j)}\leftarrow\mathbf{\Phi}^{(j)}\setminus\{\phi^{(j)}\}$
		\Until{$\exists \omega \in \Omega$ such that \red{$\hat{E}^{(j)}_{\mathbf{\Phi}}(\im\omega)\notin \mathcal{E}^{(j)}(\omega)$}}
\end{algorithmic}
\textbf{Output:} $\mathbf{\Phi}^{(j)}\cup\{\phi^{(j)}\}$
\end{algorithm}
%\vspace{.1cm}
\end{minipage} 
\begin{minipage}{.5\textwidth}
\centering
\begin{algorithm}[H]
\caption{RMI-R Incremental selection}\label{alg:RMIsub2}
\textbf{Input: } $\Sigma^{(j)}, \ \mathbf{\Phi}_\epsilon^{(j)}, \ \E^{(j)}(\omega)$ for $\omega \in \Omega$
\begin{algorithmic}[1]
		\State $\mathbf{\Phi}^{(j)} = \mathbf{\Phi}_\epsilon^{(j)}$
		\Repeat
		\State $\phi^{(j)}\leftarrow\argmin\limits_{\phi^{(j)}\in\mathbf{\Phi}^{(j)}} \operatorname{RMI-R}(\mathbf{\Phi}^{(j)},\phi^{(j)})$
		\State $\mathbf{\Phi}^{(j)}\leftarrow\mathbf{\Phi}^{(j)}\setminus\{\phi^{(j)}\}$
		\Until{$\exists \omega \in \Omega$ such that \red{$\hat{E}^{(j)}_{\mathbf{\Phi}}(\im\omega)\notin \mathcal{E}^{(j)}(\omega)$}}
\end{algorithmic}
\textbf{Output:} $\mathbf{\Phi}^{(j)}\cup\{\phi^{(j)}\}$
\end{algorithm}
%\vspace{.1cm}
\end{minipage}\\
\hspace{.2cm}
\begin{minipage}{.5\textwidth}
	\centering
   	\includegraphics[scale=.8, page=5]{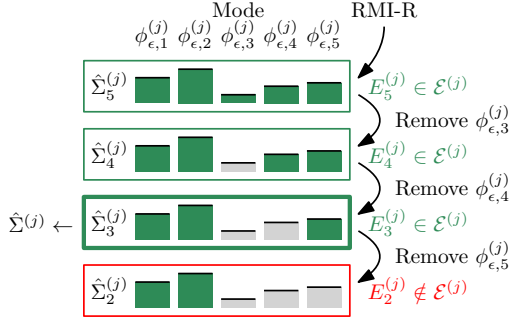}
   	\captionof{figure}{RMI-R A priori approach: Compute RMI-R (indicated by the height of the bars) of all eigenmodes initially, remove eigenmodes until \red{$E^{(j)}\notin\E^{(j)}$}.}
    \label{fig:sub_aprio}
\end{minipage} 
\hspace{.5cm}
\begin{minipage}{.5\textwidth}
	\centering
   	\includegraphics[scale=.8, page=6]{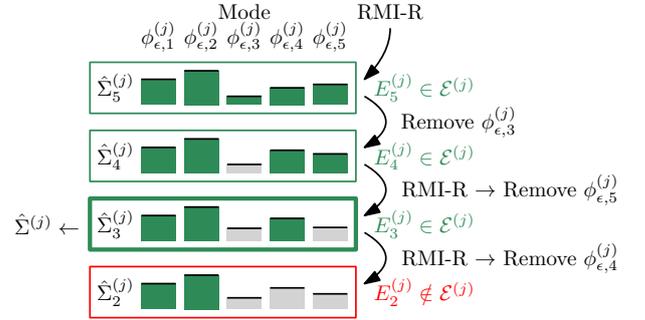}
   	\captionof{figure}{RMI-R Incremental approach: Compute RMI-R of all eigenmodes at every iteration, remove eigenmodes until \red{$E^{(j)}\notin\E^{(j)}$}.}
    \label{fig:sub_greedy}
\end{minipage}\hspace{-.8cm}

\begin{figure}
  	\centering
   	\includegraphics[scale=.7, page=1]{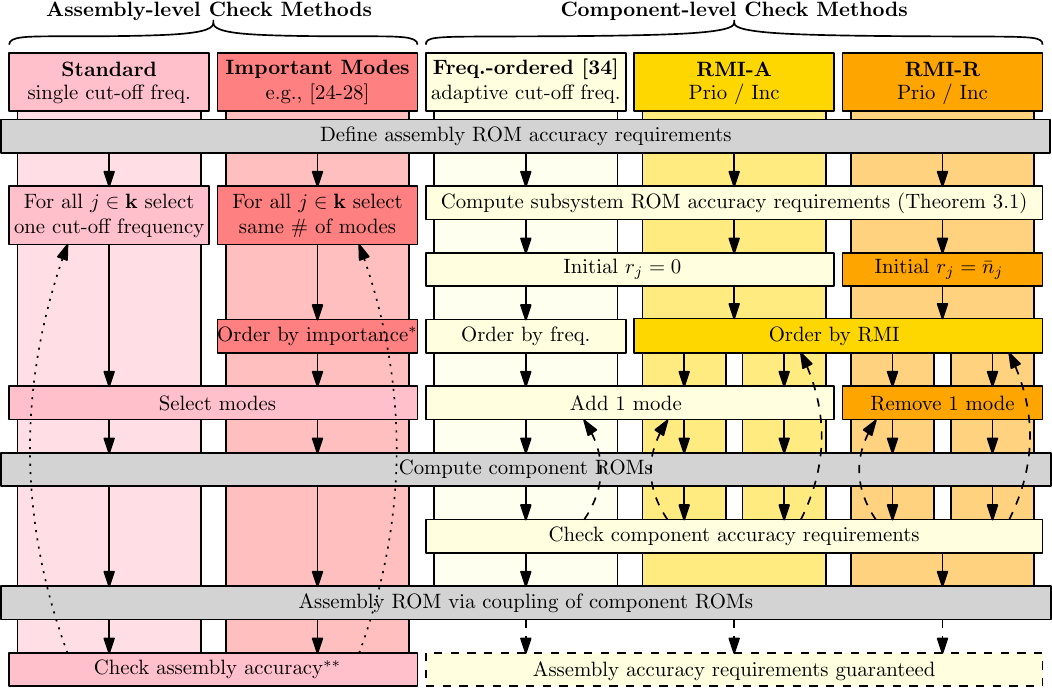}
   	\caption{Component eigenmode selection methods overview. $^*$Selection of most important modes differs per method. $^{**}$Due to computational complexity, checking the assembly requirements is often not feasible in practice. For both the RMI-A and RMI-R methods, the left dashed arrows indicate the \emph{A priori} (Prio) variant and the right dashed arrows the \emph{Incremental} (Inc) variant.}
    \label{fig:mode_selection}
\end{figure}

An overview of different component eigenmode selection methods, including the standard global method (in which the same cut-off frequency is used for all components), several other global eigenmode selection methods, the frequency-ordered method from~\cite{janssen2023translating}, and the four new methods, is given schematically in Figure~\ref{fig:mode_selection}.

\subsection{Method comparison: Computational cost}
In this subsection, the computational complexity of the mode selection algorithms is estimated, analysed, and discussed.
The computational complexity of selecting the minimum number of elastic free-interface eigenmodes of component $j$ to fulfill the component ROM accuracy requirements relies on 1) the number of iterations required to obtain the final selection of eigenmodes and 2) the computational cost of these iterations.
 
\begin{figure}
  	\centering
   	\includegraphics[scale=.6]{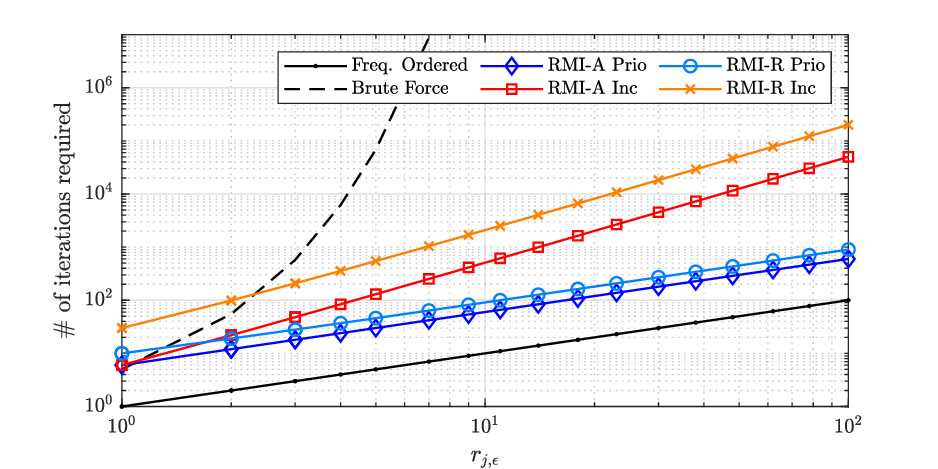}
   	\caption{Comparison of the number of iterations required for each method with respect to the number of selected eigenmodes $r_{j,\epsilon}$ given $\bar{n}_{j,\epsilon} = 5 r_{j,\epsilon}$.}
    \label{fig:cost}
\end{figure}

\begin{enumerate}
\item The number of iterations required to select the eigenmodes for the different methods, ordered by number of required iterations, is given in Table~\ref{tab:its}.
%In this table, its is assumed that $r_{j,\epsilon} \ll \bar{n}_{j,\epsilon}$. 
Recall that $\bar{n}_{j,\epsilon} \leq n_{j,\epsilon}$ is the total number of preselected eigenmodes considered in the algorithm, i.e. the number of eigenmodes left after the preselection based on a cut-off frequency, and $r_{j,\epsilon}$ is the final number of selected eigenmodes.
To reduce the required number of iterations, a lower $\bar{n}_{j,\epsilon} < n_{j,\epsilon}$ can be picked for the preselection, e.g., all eigenmodes up to five times the maximum frequency of interest.
In Figure~\ref{fig:cost}, as an illustration, the total number of required iterations is illustrated for the different mode selection methods with respect to the number selected modes $r_{j,\epsilon}$ and given $\bar{n}_{j,\epsilon} = 5 r_{j,\epsilon}$.
%\begin{remark}
%Note that there are several different variations of the four proposed mode selection methods that could lead to a lower number of iterations required or better results, e.g.,
%\begin{enumerate}
%\item both for the RMI-A and RMI-R approaches, instead of updating the RMI either a priori or at every iteration (incremental), the RMI values can be updated at a specific interval, e.g., every 5 iterations.
%By doing so, the advantages of the a priori approach and the incremental approach can be combined: a reduced number of iterations is required for the a priori approach which is combined with the advantage of the incremental approach that the RMI values are updated during the selection approach.
%\item Alternatively, for the a priori approach, the RMI values can be computed using RMI-R a priori (Algorithm~\ref{alg:RMIsub}) approach. 
%Then, use the RMI-A approach (Algorithm~\ref{alg:RMIadditive}) to select the eigenmodes with the obtained RMI-R values instead of the RMI-A values.
%This will automatically result in the eigenmodes that would normally be selected by the RMI-R a priori approach but with the number of iterations required for the RMI-A a priori approach.
%Therefore, if $r_{j,\epsilon} < \frac{1}{2}{\bar{n}_{j,\epsilon}}$, this would reduce the number of iterations required. 
%\end{enumerate}
%However, to keep this work as concise as possible, these mixed approaches are not evaluated explicitly in the remainder of the paper.
%\end{remark}
\item The computational cost of each iteration is relatively low, because the approach is on component level.
Namely, to check if a specific selection of component eigenmodes satisfies the assembly requirements $E_A(\im\omega)\in\E_A(\omega)$ for any $\omega \in \Omega$, only the component requirements $E^{(j)}(\im\omega) \in \E^{(j)}(\omega)$ need to be checked.
To determine $E^{(j)}(\im\omega) = \hat{H}^{(j)}(\im\omega)-H^{(j)}(\im\omega)$, the high-order component FRF $H^{(j)}(\im\omega)$ has to be determined only once.
Therefore, only the component ROM FRF $\hat{H}^{(j)}(\im\omega)$ has to be determined for every iteration, which, depending on the system, as mentioned in Section~\ref{sec:framework}, can often be done effectively using modal analysis.
\end{enumerate}
\begin{table}[]
\caption{Comparison of the number of iterations required for each selection method and the order of magnitude assuming $r_{j,\epsilon} \ll \bar{n}_{j,\epsilon}$. }
\label{tab:its}
\begin{tabular}{l|ll}
\textbf{Method} & \textbf{\# of iterations} & \textbf{Order of magitude} (for $r_{j,\epsilon} \ll \bar{n}_{j,\epsilon}$)\\
\hline
Freq. Ordered & $r_{j,\epsilon}$ & $\mathcal{O}(r_{j,\epsilon})$ \\
RMI-A A Priori & $ \bar{n}_{j,\epsilon} + r_{j,\epsilon}$&  $\mathcal{O}(\bar{n}_{j,\epsilon})$  \\
RMI-R A Priori & $2\bar{n}_{j,\epsilon} - r_{j,\epsilon}$& $\mathcal{O}(\bar{n}_{j,\epsilon} \times 2)$ \\
RMI-A Incremental & $ (\bar{n}_{j,\epsilon} + 1) r_{j,\epsilon}$& $\mathcal{O}(\bar{n}_{j,\epsilon} \times r_{j,\epsilon})$ \\
RMI-R Incremental & $(\bar{n}_{j,\epsilon} + 1)(\bar{n}_{j,\epsilon}-r_{j,\epsilon})$& $\mathcal{O}(\bar{n}_{j,\epsilon} ^2)$ \\
Brute Force & $\sum_{q=1}^{r_{j,\epsilon}}\frac{\bar{n}_{j,\epsilon}!}{(\bar{n}_{j,\epsilon}-q)! q!}$ & $\mathcal{O}(\bar{n}_{j,\epsilon}^{r_{j,\epsilon}})$
\end{tabular}
\end{table}

\section{ILLUSTRATIVE CASE STUDIES}
\label{sec:examples}
In this section, the mode selection methods proposed in Section~\ref{sec:mode_selection} are demonstrated on two example systems.
Here, we recall that the assembly ROM accuracy requirement $\E_A$ is free to be designed by the user.
However, for both systems, we consider a specific assembly ROM accuracy requirement $\E_A$ based on a maximum on the relative error, given by $\gamma$, i.e., 
\begin{align}
\label{eq:ex_req}
\frac{\|E_A(\im\omega)\|}{\|H_A(\im\omega)\|} = \frac{\|\hat{H}_A(\im\omega) - H_A(\im\omega)\|}{\|H_A(\im\omega)\|} < \gamma,
\end{align}
for all $\omega \in \hat{\Omega}$ where $\hat{\Omega}$ is a discrete subset of $\Omega$. 

\subsection{Coupled cantilever beams}
\begin{figure}
\begin{subfigure}{.3\textwidth}
  	\centering
   	\includegraphics[scale=1, page = 1]{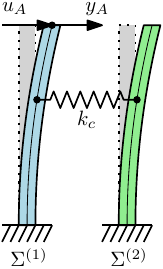}
    \caption{Scematic drawing}
    \label{fig:example_1}
\end{subfigure}
\begin{subfigure}{.7\textwidth}
  	\centering
   	\includegraphics[scale=.6]{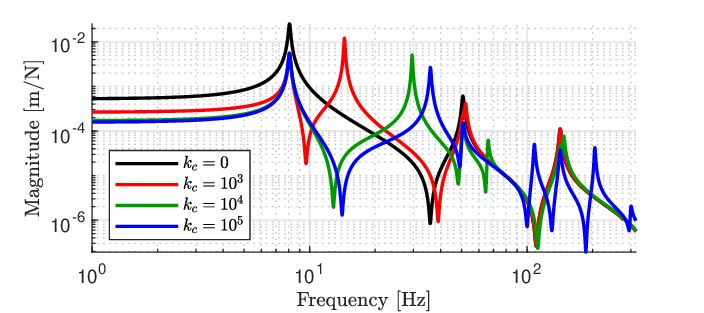}
   	\caption{External input($u_A$)-output($y_A$) FRFs $|H_A(\im\omega)|$ for different values of $k_c$.}
   	\label{fig:cantilever_FRF}
\end{subfigure}
\caption{Two cantilever beams interconnected by a spring with stiffness $k_c$ [N/s]. The system has an external input force of $u_A$ [N] and an external output displacement of $y_A$ [m].}
\end{figure}

The first case study is a simple academic mechanical system of two interconnected cantilever beams, in which the value of the coupling spring stiffness is varied to illustrate the approaches described in Section~\ref{sec:methodology} and to show how the different mode selection methods from Section~\ref{sec:mode_selection} compare. 
The system is schematically shown in Figure~\ref{fig:example_1}.

For building the high-order component models $\Sigma^{(1)}$ and $\Sigma^{(2)}$, each cantilever beam is discretized by 50 linear two-node Euler beam elements (only bending, no shear, see~\cite{craig2006}) of equal length. 
Per node we have one (transversal) translational and one rotational DOF ($n_1 = n_2 = 100$), i.e., we consider bending of the beams in the plane of Figure~\ref{fig:example_1}.
Furthermore, the components both have a length of $1$ m, a cross-sectional area of $10^{-4}$ m$^2$, a 2nd area moment of $1/12\times 10^{-8}$ m$^4$, a Young's modulus of $2\times 10^{11}$ Pa and a mass density of $8\times 10^{3}$ kg/m$^3$.
Viscous damping of both components is modelled using 1\% modal damping.
%The rotation and translation of the 1st DOF is constrained (to the fixed world) for both components.

The left beam ($\Sigma^{(1)}$) has an external input force $u_A$ at its free end and the collocated transversal displacement is measured via the external output $y_A$.
Furthermore, between the 33rd translational DOFs of both components, a translational spring with a stiffness $k_c$ [N/m] is positioned, which interconnects the components.
The interconnection structure of the system is given by
\begin{align}
\left[\begin{array}{c}
u^{(1)}_1 \\ 
u^{(1)}_2 \\ \hline
u^{(2)}_1 \\ \hline
y_A
\end{array}\right] = \K
\left[\begin{array}{c}
y^{(1)}_1 \\ 
y^{(1)}_2 \\ \hline
y^{(2)}_1 \\ \hline
u_A
\end{array}\right] \text{ with } \K = \left[\begin{array}{cc|c|c}
0 & 0 & 0& 1\\
0 & -k_c & k_c& 0\\ \hline
0 & k_c & -k_c& 0\\ \hline
1 & 0 & 0& 0
\end{array}\right].
\end{align}
%Here, $y^{(j)} = \Sigma^{(j)}u^{(j)}$ for $j = 1,2$ and $y_c = \Sigma_c u_c$.
The FRFs for different values of $k_c$ are given in Figure~\ref{fig:cantilever_FRF}. Note that for $k_c = 0$, $H_A(\im\omega) = H^{(1)}_1(\im\omega)$, as there is no interaction with $\Sigma^{(2)}$.
In this case study, the discrete frequency range of interest $\hat{\Omega}$ is defined by taking $100$ logarithmically distributed frequency points from $0.1$ Hz up to the maximum frequency of interest $f_{\max} := \omega_{\max} / (2\pi) = 400$ Hz.
Furthermore, a relative error of $\gamma = 0.05$ is the required ROM assembly accuracy, see (\ref{eq:ex_req}).
To describe (\ref{eq:ex_req}) within the accuracy requirement framework (\ref{eq:Ec_bound}), the weighting matrices $V_A(\omega)$ and $W_A(\omega)$ are chosen as $V_A(\omega) = W_A(\omega) =(\gamma\|H_A(\im\omega)\|)^{-\frac{1}{2}}$.
Then, with $V_A(\omega)$ and $W_A(\omega)$, the optimization problem (\ref{eq:top_down}) can be solved using the approach described in Section~\ref{sec:methodology} for all $\omega \in \hat{\Omega}$.
The resulting (unique) solution consists of $V^{(j)}(\omega)$ and $W^{(j)}(\omega)$ and, therefore, component ROM accuracy requirements $\E^{(j)}(\omega)$ are obtained for all $j \in \bk$.

\begin{figure}
  	\centering
   	\includegraphics[scale=.6]{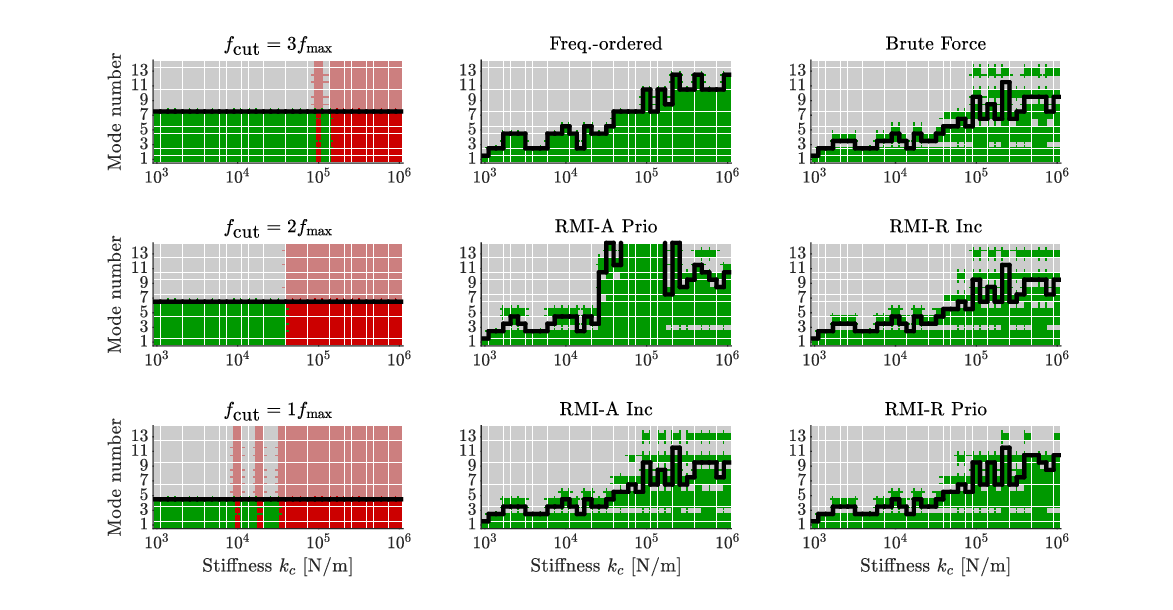}
   	\caption{Eigenmode selection method comparison for varying interconnection stiffness $k_c$. \legendbox{green_2} Mode selection for which \green{$E_A\in\E_A$}, \legendbox{red} Mode selection for which \red{$E_A\notin\E_A$}, \legendbox{grey_1} Not selected modes, \legendline{black} total number of eigenmodes $n_{2,\epsilon}$.}
    \label{fig:cantilever_stiffness}
\end{figure}

\begin{figure}
\begin{subfigure}{.5\textwidth}
  	\centering
   	\includegraphics[scale=.6]{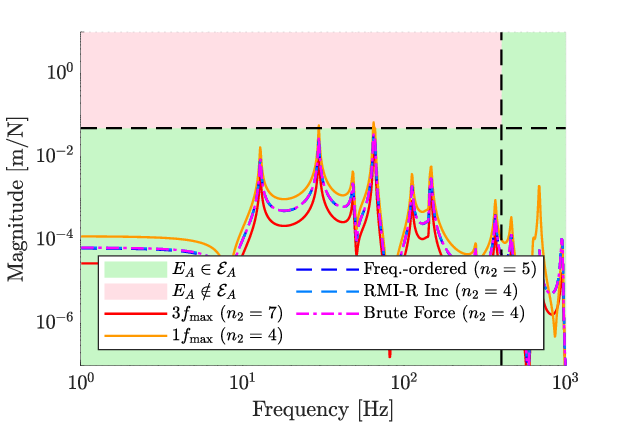}
    \caption{Results for $k_c = 10^{4}$ N/m. Only the assembly ROM obtained using $f_{\textrm{cut}} = 1f_{\max}$ does not satisfy the requirement, i.e., \red{$E_A\notin\E_A$}. The other ROMs satisfy the requirement, i.e., \green{$E_A\in\E_A$}.}
    \label{fig:FRF_low}
\end{subfigure}
\begin{subfigure}{.5\textwidth}
  	\centering
   	\includegraphics[scale=.6]{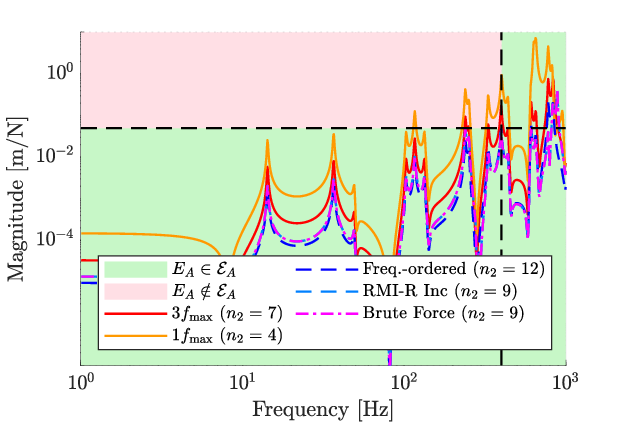}
    \caption{Results for $k_c = 10^{6}$ N/m. The assembly ROMs obtained using $f_{\textrm{cut}} = 1f_{\max}$ and $f_{\textrm{cut}} = 3f_{\max}$ do not satisfy the requirement, i.e., \red{$E_A\notin\E_A$}. The other ROMs satisfy requirement, i.e., \green{$E_A\in\E_A$}.}
    \label{fig:FRF_high}
\end{subfigure}
   	\caption{Relative errors $\|E_A(\im\omega)\|/\|H_A(\im\omega)\|$ of external input-output behavior of the cantilever system assembly ROMs $\hat{\Sigma}_A$ obtained with different mode selection methods, and the assembly ROM accuracy requirement $\E_A$, for two different values of $k_c$.}
   	\label{fig:cantilever_result_FRF}
\end{figure}

In this example, the first component is not reduced, i.e., $\hat{\Sigma}^{(1)} = \Sigma^{(1)}$.
%Therefore, for the assembly accuracy based mode selection methods, only the component ROM accuracy requirements $\E^{(2)}$ are computed using (\ref{eq:top_down}).
The proposed mode selection methods in Section~\ref{sec:mode_selection} are applied to the second component for different values of $k_c$ to compute component ROM $\hat{\Sigma}^{(2)}$.
In addition, the standard CMS approach is applied to component 2 with a cut-off frequency of $1,2$ and $3$ times the maximum frequency of interest $f_{\max}$, i.e., $400, 800$ and $1\,200$ Hz.
In Figure~\ref{fig:cantilever_stiffness}, the results of these analyses are given.
In addition, the relative errors $\|E_A(\im\omega)\|/\|H_A(\im\omega)\|$ obtained for two stiffness values ($k_c = 10^4$ N/m and $k_c = 10^6$ N/m) for five of the compared mode selection methods are given in Figure\ref{fig:cantilever_result_FRF}. Note that this figure is obtained a posteriori, by computing the FRF of the complete high-order assembly model $\Sigma_A$.

By definition, the standard approach selects a number of component eigenmodes independent of the assembly or the interaction between components. 
Therefore, the number of selected component eigenmodes remains equal if the interconnection matrix $\K$ changes.
In contrast, in the approaches based on assembly accuracy requirements, the obtained component ROM accuracy requirement $\E^{(2)}$ takes into account the influence of input-output errors in $\Sigma^{(2)}$ to the assembly ROM accuracy requirement $\E_A$.
As a result, automatically, if $\K$ changes, so does the component ROM accuracy requirement $\E^{(2)}$, and consequently the required number of selected component eigenmodes.
For example, for a low coupling stiffness $k_c$, fewer component eigenmodes of component 2 in general are required to satisfy $\E_A$ because the second cantilever beam ($\Sigma^{(2)}$) has less influence on the assembly, in terms of external input($u_A$)-to-output($y_A$) behavior, when the stiffness of the interconnecting spring is low.

%For the results in Figure~\ref{fig:cantilever_stiffness}, the average CPU time required obtain a selection of $r_2$ component eigenmodes from $\bar{n}_2 = 30$ for all components for all stiffness values is $0.02$ s for the Freq.-ordered approach, $0.09$ s for the RMI-A A priori approach $0.26$ s for the RMI-A Incremental approach, $0.07$ s for the RMI-R A priori approach, $0.46$ s for the RMI-R Incremental approach, and $11.23$ s for the Brute Force approach. Please note that the computations in this work are applied using on a modern notebook (Intel i7 2.6Ghz Processor, 16Gb RAM) and that CPU times reported in this work should be viewed indicative performance \emph{relative} to each other as the absolute values are obviously subject to the computing power available. 

When comparing the results of the different component eigenmode selection methods for this case study in Figure~\ref{fig:cantilever_stiffness}, it is clear that the frequency-ordered method from \cite{janssen2023translating} generally cannot achieve the optimal result which is achieved by the brute force method, as it can only select the component eigenmodes in order of increasing eigenfrequency. 
Furthermore, the RMI-A a priori approach does not achieve a satisfactory result, as, in general, more component eigenmodes are required compared to the frequency-ordered method.
%For this example, this mode selection method is therefore not effectively selecting appropriate modes.
However, the other RMI-based selection methods, i.e., the Incremental approaches, although they selected different component eigenmodes, can achieve an $r_2$ \emph{equal to the optimal} number obtained using the brute force approach for all tested values of $k_c$ at a significantly lower computational cost.
Finally, the RMI-R a priori achieves a selection of eigenmodes which for some values of $k_c$ requires (only) one eigenmode extra with respect to the optimal solution.

This simple case study already shows that, because the industrial standard approach does not take into account the interconnection structure, none of the three choices of the cut-off frequency are able to select a proper amount of component eigenmodes for all values of $k_c$, i.e., sometimes the assembly ROM accuracy requirements are not met (red columns), and sometimes the number of eigenmode component modes is chosen much too large (see $f_{\cut}= 3f_{\max}$ for low interconnection stiffness).
In contrast, the mode selection methods based on assembly accuracy obtain, in a computationally efficient manner, an assembly ROM which accounts for the interaction between components and therefore selects an appropriate amount of modes the component.
Note, however, that in the case of this simple beam system, the number of (component) eigenmodes in the frequency range of interest is low, and that the full potential of the new component eigenmode selection methods is better demonstrated in the case of more complex systems with a higher number of (component) eigenmodes in the frequency range of interest.
Therefore, in the next section, we will use a more complex case study with a higher number of (component) eigenmodes in the frequency range of interest.

\subsection{2D Wirebonder model}
\begin{figure}
\begin{subfigure}{.5\textwidth}
  	\centering
   	\includegraphics[scale=0.18]{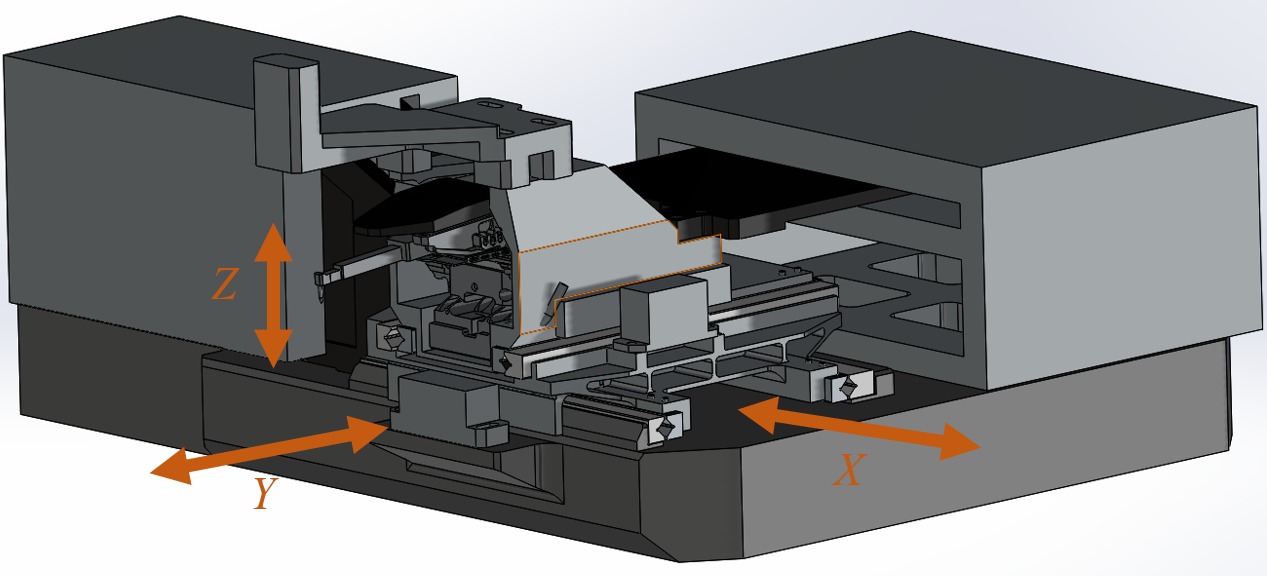}
%   	\vspace{1pt}
   	\caption{Schematic picture of a 3D Industrial wire bonder system. X, Y and Z indicate the directions in which the 3 stages move.}
    \label{fig:wirebonder_pic}
\end{subfigure}
\label{fig:wirebonder}
\begin{subfigure}{.5\textwidth}
  	\centering
   	\includegraphics[trim={.8cm 1cm 1.6cm 2.2cm},clip, scale=0.55]{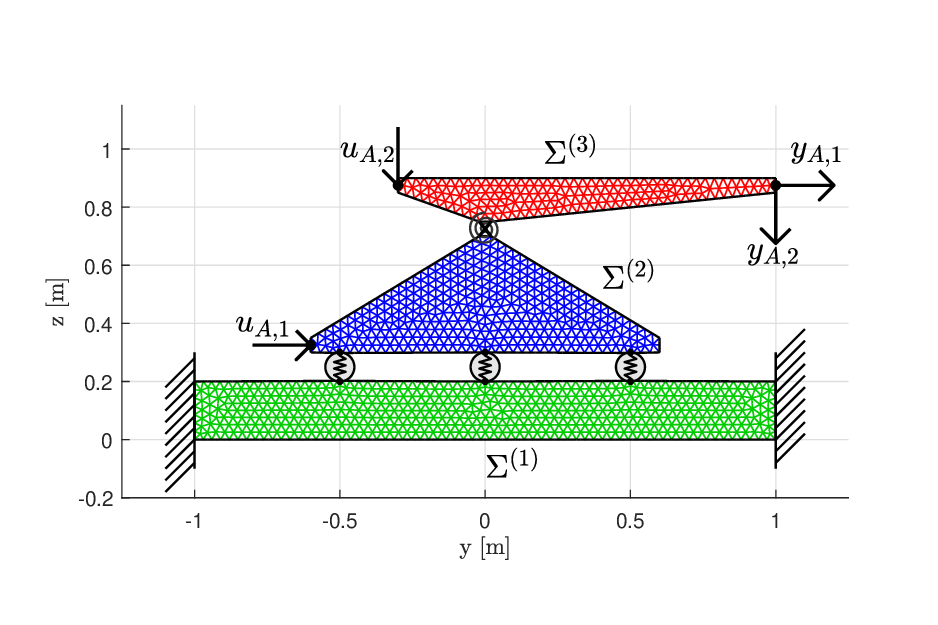}
   	\caption{2D Finite Element model with interconnections between fixed x-stage ($\Sigma^{(1)}$), y-stage ($\Sigma^{(2)}$) and z-stage ($\Sigma^{(3)}$).}
    \label{fig:wirebonder_example}
\end{subfigure}
\begin{subfigure}{1\textwidth}
  	\centering
   	\includegraphics[scale=.6]{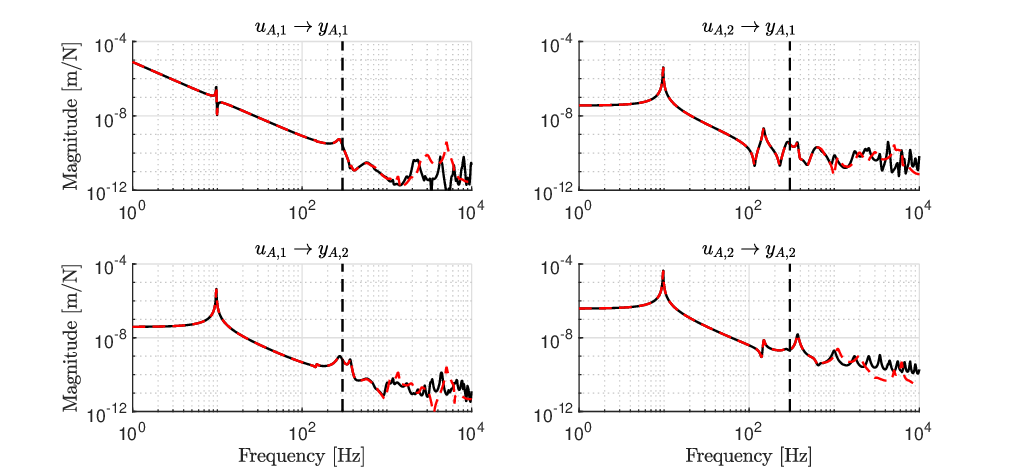}
    \caption{FRF of all external input-output pairs of the 2D wirebonder assembly model $\Sigma_A$ (black) and an example assembly ROM $\hat{\Sigma}_A$ (red) that is accurate up to $f_{\max} = 300$ Hz.}
    \label{fig:FRF_WB}
\end{subfigure}
\caption{Industrial and simplified model of a wirebonder machine.}
\end{figure}

In the case study covered in this section, we apply the proposed approach on the 2D structural dynamics FE assembly model that was used in \cite{janssen2023translating}.
The model is inspired by a 3D industrial wire bonder system (Figure~\ref{fig:wirebonder_pic}). 
The FE assembly model consists of three elastically interconnected components (Figure~\ref{fig:wirebonder_example}) which are simplifications of the x-, y-, and z-stage of the industrial model, and where the x-stage is connected to the fixed world.
The system in Figure~\ref{fig:wirebonder_example} allows for rigid body translation of the y-stage ($\Sigma^{(2)}$) relative to the fixed x-stage ($\Sigma^{(1)}$) and rotational motion of the z-stage ($\Sigma^{(3)}$) about the y-stage via a cross leaf-spring. 
To demonstrate our methodology, we will linearize the model around one working point.
The elaborated model specifications and the chosen parameter values can be found in \cite{janssen2023translating}.
For this system, the coupling of the three components, as generally described by (\ref{eq:sigmac}), is given by
\begin{align}
\left[\begin{array}{c}
u^{(1)} \\ \hline
u^{(2)} \\ \hline
u^{(3)} \\ \hline
y_{A,1} \\
y_{A,2} 
\end{array}\right] = \left[\begin{array}{c|cc|ccc|cc}
{-}K_b & K_b & & & & & & \\
\hline
K_b & {-}K_b & & & & & & \\
& & & & & & 1 & \\
& & {-}K_{cl} & K_{cl} & & & & \\
\hline
& & K_{cl}& {-}K_{cl} & & & & \\
& & & & & & & 1 \\
\hline
& & & & 1 & & & \\
& & & & & 1 & & \\
\end{array}\right]
\left[\begin{array}{c}
y^{(1)} \\ \hline
y^{(2)} \\ \hline
y^{(3)} \\ \hline
u_{A,1} \\
u_{A,2} 
\end{array}\right].
\end{align}
Here, $K_b = \text{diag}(k_b, k_b, k_b)$ and $K_{cl} = \text{diag}(k_t, k_t, k_r)$ where $k_b$ [N/m] is both the vertical bearing stiffness, $k_t$ [N/m] is the horizontal and the vertical cross leaf-spring stiffness and $k_r$ [Nm/rad] the rotational cross leaf-spring stiffness.
Furtermore, note that $u^{(1)} \in \R^3$, $u^{(2)} \in \R^7$, $u^{(3)}  \in \R^4$, $y^{(1)}  \in \R^3$, $y^{(2)}  \in \R^6$, and $y^{(3)}  \in \R^5$.
Then, assembly ROM accuracy requirements for all four FRFs from $u_{A,1}$ and $u_{A,2}$ to $y_{A,1}$ and $y_{A,2}$ can be defined.
The FRF of the assembly model $\Sigma_A$ and an example assembly ROM $\hat{\Sigma}_A$ are given in Figure~\ref{fig:FRF_WB}.

In this case study, $\hat{\Omega}$ is defined by the set of $500$ logarithmically distributed frequency points from $0.1$ Hz up to the maximum frequency of interest $f_{\max}$.
Furthermore, the maximum frequency of interest $f_{\max}$ is varied to show how the component eigenmode selection methods perform for assembly ROMs with a relatively low to a higher frequency range of interest in comparison to the standard CMS cut-off frequency approach and to each other.
A relative error of $\gamma = 0.01$ in (\ref{eq:ex_req}), i.e., 1\% relative error, is the required assembly ROM accuracy for all possible external input-to-output combinations of the system for all $\omega \in \hat{\Omega}$. 
To describe (\ref{eq:ex_req}) within the accuracy requirement framework (\ref{eq:Ec_bound}), the weighting matrices $V_A(\omega)$ and $W_A(\omega)$ are chosen as $V_A(\omega) = W_A(\omega) =\frac{1}{2}I_2(\gamma\|H_A(\im\omega)\|)^{-\frac{1}{2}}$.
For several different maximum frequencies of interest $f_{\max}$, the resulting total number of component eigenmodes $r_{A,\epsilon}$, and the CPU time to obtain these results, are given in Table~\ref{tab:n_c}.  
Please note that the computations in this work are applied using on a modern notebook (Intel i7 2.6Ghz Processor, 16Gb RAM) and that CPU times reported in this work should be viewed indicative performance \emph{relative} to each other as the absolute values are obviously subject to the computing power available. 
%\begin{remark}
%\label{rem:cut-off-time}
%Since we the compare the eigenmode selection methods in this work, we compare only the CPU time required to obtain a selection of $r_j$ component eigenmodes from $\bar{n}_j$.
%In comparison, no computational effort is required for the cut-off frequency approaches, as $r_j$ is selected a priori.
%However, with the cut-off frequency approaches, no guarantees on the accuracy of assembly ROM are available.
%In fact, in Figures~\ref{fig:cantilever_stiffness} and \ref{fig:cantilever_result_FRF}, it can be seen that the requirements are not (always) satisfied using the cut-off frequency approach.
%Therefore, to obtain such guarantees, the obtained assembly ROM accuracy would need to be compared with the high-order assembly model $\Sigma_A$.
%Recall this computationally expensive task is avoided for the RMI-based, frequency-ordered, and the brute force approaches as satisfying the component ROM accuracy requirements $\E^{(j)}$ automatically guarantees the satisfaction of assembly ROM accuracy requirements $\E_A$.
%\end{remark}
\begin{table}[]
\centering
\caption{Total number of component eigenmodes $r_{A,\epsilon} = \sum_{j=1}^3r_{j,\epsilon}$ selected in the assembly ROM $\hat{\Sigma}_A$ of the 2D wirebonder FE model (top part of the table) and the required CPU time to obtain these results for the modular approaches in seconds (bottom part of the table). The CPU time is the total time required obtain a selection of $r_{j,\epsilon}$ component eigenmodes from $\bar{n}_{j,\epsilon}$ for all components. $^*$For $f_{\cut} = 1f_{\max}$ and for some values of $f_{\max}$ when $f_{\cut} = 2f_{\max}$, the assembly  ROM accuracy requirements are not satisfied, i.e., \red{$E_A \notin \E_A$}. $^{**}$The brute force method becomes ``infeasible'', i.e., the estimated CPU time is $>10^9$ ($>30$ years) seconds. The best result(s), i.e., the assembly ROM with the smallest $n_{A,\epsilon}$ that satisfies \green{$E_A \in \E_A$}, is/are given in \green{\textbf{bold}}.}
\label{tab:n_c}
\begin{tabular}{l|lllllll}
$f_{\max}$ [Hz]        & $300$                & $600$                & $1\,200$              & $3\,000$               & $6\,000$               & $12\,000$\\ \hline
\textbf{\# of eigenmodes} $r_{A,\epsilon}$ & & & & & & \\
$f_{\cut} = 3f_{\max}$ & \green{3}          & \green{9}          & \green{18}         & \green{52}          & \green{156}         & \green{512} \\
$f_{\cut} = 2f_{\max}$ & \green{\textbf{2}} & \red{5$^*$}        & \green{11}         & \green{30}          & \green{76}          & \green{256} \\
$f_{\cut} = 1f_{\max}$ & \red{1$^*$}        & \red{2$^*$}        & \red{5$^*$}        & \red{15$^*$}        & \red{30$^*$}        & \red{76$^*$} \\ 
Freq. Ordered          &  \green{\textbf{2}} & \green{\textbf{5}} & \green{\textbf{7}} & \green{20}          & \green{36}          & \green{85} \\
RMI-A A priori      & \green{\textbf{2}} & \green{\textbf{5}} & \green{\textbf{7}} & \green{23}          & \green{39}          & \green{105} \\
RMI-A Incremental        & \green{\textbf{2}} & \green{\textbf{5}} & \green{\textbf{7}} & \green{\textbf{16}} & \green{\textbf{30}} & \green{72} \\
RMI-R A priori      & \green{\textbf{2}} & \green{\textbf{5}} & \green{\textbf{7}} & \green{\textbf{16}} & \green{\textbf{30}} & \green{91} \\
RMI-R Incremental        & \green{\textbf{2}} & \green{\textbf{5}} & \green{\textbf{7}} & \green{\textbf{16}} & \green{\textbf{30}} & \green{\textbf{68}} \\ 
Brute Force         & \green{\textbf{2}} & \green{\textbf{5}} & \green{\textbf{7}} & \green{\textbf{16}} & \green{\textbf{30}} & ?$^{**}$ \\ \hline
\textbf{CPU time} [s] & & & & & & & \\
Freq. Ordered       &   0.02 &   0.02 &   0.03 &   0.07 &   0.12 &   0.39 \\
RMI-A A priori      &   0.03 &   0.06 &   0.07 &   0.19 &   0.32 &   0.97 \\
RMI-A Incremental        &   0.02 &   0.04 &   0.07 &   0.25 &   0.87 &   6.58 \\
RMI-R A priori      &   0.02 &   0.04 &   0.06 &   0.18 &   0.39 &   1.70 \\
RMI-R Incremental        &   0.02 &   0.06 &   0.07 &   0.60 &   2.06 &  17.81 \\ 
Brute Force         &   0.01 &   0.03 &   0.06 &   6.31 & 4\,434.37 \hspace{-.3cm} & ?$^{**}$
\end{tabular}
\end{table}

In the results given in this table, it can been seen that the results on the 2D wirebonder system are qualitatively comparable to the results of the previous case study (coupled cantilever beams).
Namely, the assembly ROM obtained using the component eigenmode selection approaches are significantly further reduced with respect to the standard approaches where $f_{\cut} = 2f_{\max}$ and $f_{\cut} = 3f_{\max}$ are used, whilst $f_{\cut} = 1f_{\max}$ does generally not satisfy the assembly ROM accuracy requirements.

For this system, it also holds that the RMI-A A priori approach is not able to outperform the frequency-ordered approach from \cite{janssen2023translating} in terms of the required component eigenmodes, whereas the other RMI approaches achieve the optimal number of selected component eigenmodes that was also selected by the brute force method.
For $f_{\max}=12\,000$ Hz, where application of the brute force method became infeasible with an estimated CPU time of over 30 years (!), the RMI-R Incremental approach is able to achieve the lowest order assembly ROM, at the cost of a slightly longer CPU time (17.81 seconds) compared to the other RMI approaches.
%However, for this case, the brute force method was deemed infeasible, as the estimated CPU time is $>10^9$ seconds. 
Because the optimal brute force result is lacking, it cannot be verified whether this result is the optimal component eigenmode selection.
\begin{figure}
\begin{subfigure}{.5\textwidth}
  	\centering
   	\includegraphics[scale=.6]{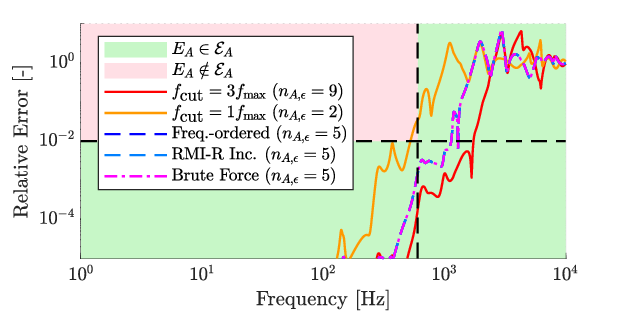}
    \caption{Results for $f_{\max} = 600$ Hz.}
    \label{fig:FRF_WB_low}
\end{subfigure}
\begin{subfigure}{.5\textwidth}
  	\centering
   	\includegraphics[scale=.6]{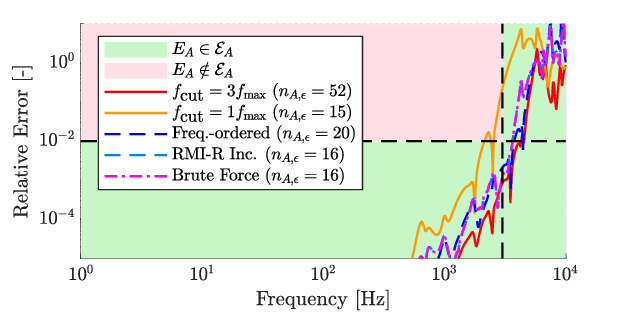}
    \caption{$f_{\max} = 3\,000$ Hz.}
    \label{fig:FRF_WB_high}
\end{subfigure}
   	\caption{Relative errors $\|E_A(\im\omega)\|/\|H_A(\im\omega)\|$ of external input-output behavior of the wirebonder assembly ROMs $\hat{\Sigma}_A$ obtained with different mode selection methods, and the assembly ROM accuracy requirement $\E_A$, for two different values of $f_{\max}$.}
   	\label{fig:cantilever_WB}
\end{figure}

\begin{figure}
  	\centering
   	\includegraphics[scale=.6]{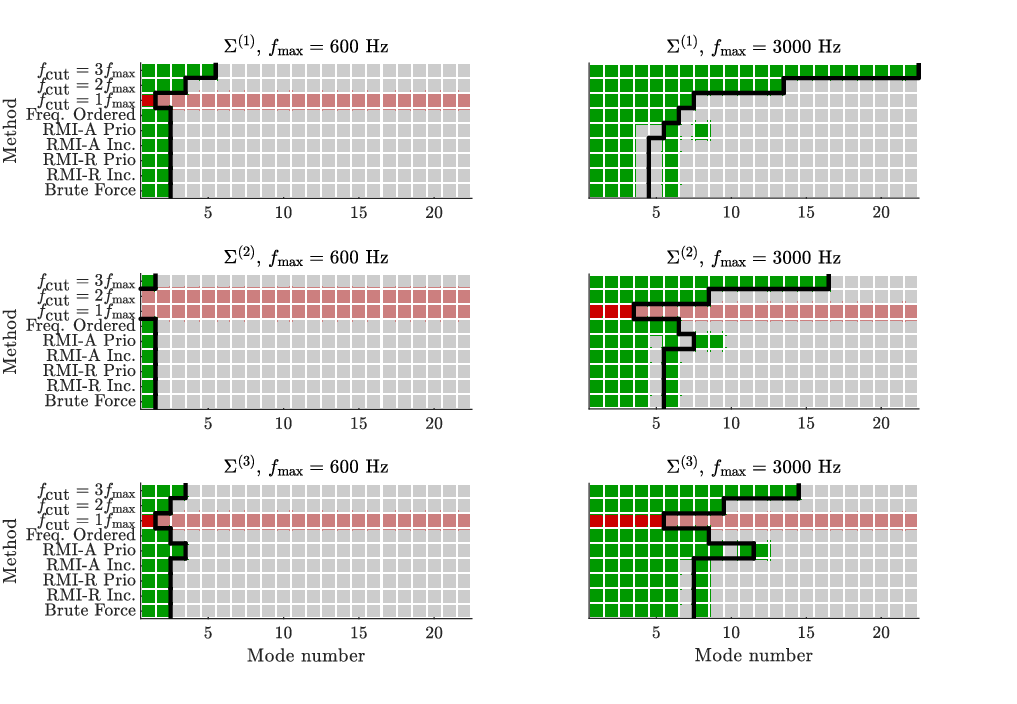}
   	\caption{Component eigenmode selection results for varying $f_{\max} = 600$ and $f_{\max} = 3\,000$ Hz for the x-, y- and z-stages of the 2D wirebonder system for the different methods. \legendbox{green_2} Mode selection for which \green{$E_A\in\E_A$}, \legendbox{red} Mode selection for which \red{$E_A\notin\E_A$} \legendline{black} total number of eigenmodes $n_{j,\epsilon}$.}
    \label{fig:freq_1p}
\end{figure}

To further illustrate these results, the relative errors $\|E_A(\im\omega)\|/\|H_A(\im\omega)\|$ obtained for two maximum frequencies of interest ($f_{\max} = 600$ Hz and $f_{\max} = 3\,000$ Hz) for five of the compared mode selection methods are given in Figure~\ref{fig:cantilever_WB}.
In these FRFs, it can be seen that only for the standard $f_{\cut} = 1f_{\max}$ approach, the assembly ROM accuracy requirement $\E_A$ is not satisfied.
Out of the other approaches, for both FRFs, the RMI-R Incremental and Brute Force approaches satisfy the assembly ROM requirements $\E_A$ with significantly fewer eigenmodes compared to the freq.-ordered approach and especially the standard $f_{\cut} = 3f_{\max}$ approach. 
Finally, to compare the \emph{individual components} results, the number of selected modes $r_{j,\epsilon}$ for each component $j=1,2,3$, two maximum frequencies of interest (again $f_{\max} = 600$ Hz and $f_{\max} = 3\,000$ Hz) are given in Figure~\ref{fig:freq_1p}.

The results are given in Figure~\ref{fig:freq_1p}. 
Because the RMI-based component eigenmode selection methods take into account the relative importance of modes to the \emph{assembly}, for the x-stage ($\Sigma^{(1)}$), a significantly lower number of eigenmodes is required to satisfy the component ROM accuracy requirements $\E^{(1)}$ compared to the number of eigenmodes which are required to satisfy the component ROM accuracy requirements for the y- and z-stage.
For example, for $f_{\max} = 3\,000$ Hz, for the RMI-R Incremental approach, $r_{1,\epsilon} = 4$ while for $f_{\cut} = 2f_{\max}$, $r_{1,\epsilon} = 13$.
In addition, for $f_{\max} = 600$, both the $f_{\cut} = 2f_{\max}$ and the proposed methods select a total of $n_{A,\epsilon} = 5$ modes. However, the number of eigenmodes \emph{per component} is different between these approaches. 
As a consequence, the $f_{\cut} = 2f_{\max}$ approach does not satisfy the required accuracy in the assembly model, as can also be seen in Table~\ref{tab:n_c}.

This illustrates how an approach for component eigenmode selection based on assembly accuracy requirements, which takes into account the relative importance of dynamics of components, can achieve component ROMs that consist of significantly less component eigenmodes compared to the standard cut-off frequency based method while guaranteeing the required accuracy of the assembly ROM.

%\begin{figure}
%  	\centering
%   	\includegraphics[scale=.8]{figures/wirebonder_freq.eps}
%   	\caption{Component eigenmode selection results for varying $f_{\max}$ for the x-, y- and z-stages of the 2D wirebonder system for the standard $f_{\cut} = 2f_{\max}$, the Frequency-ordered, the RMI-R Greedy and the Brute Force approaches. \legendbox{green_2} Mode selection for which \green{$E_A\in\E_A$}, \legendbox{red} Mode selection for which \red{$E_A\notin\E_A$}, \legendbox{orange} Mode selection infeasible (CPU time $\gg 1$ day), \legendline{black} total number of eigenmodes $n_{j,\epsilon}$.}
%    \label{fig:freq}
%\end{figure}

\section{CONCLUSIONS}
\label{sec:conclusions}
In this paper, we have addressed the challenge of reducing the complexity of (linear) high-dimensional interconnected component models while ensuring accurate representation of the dynamic behavior of the coupled, reduced assembly model within the context of CMS methods.
By translating the assembly ROM accuracy requirements to component ROM accuracy requirements using a top-down robust performance analysis framework, the most important eigenmodes for each component can be selected in a modular approach.
We have introduced the notion of relative mode importance to quantify the most important component eigenmodes given frequency-dependent component ROM accuracy requirements.
This approach based on assembly-accuracy requirements takes into account this relative importance of eigenmodes of components to the assembly dynamics, allowing for deleting unnecessary eigenmodes and the inclusion of crucial eigenmodes.
By doing so, this approach avoids the limitations of the traditional industry approach within CMS techniques, which selects component eigenmodes based on a simple predefined cut-off frequency for all components. 

We demonstrated the effectiveness of the approach through two case-studies. 
The results showed that the proposed modular and assembly accuracy based component eigenmode selection methods outperformed the industrial standard approach both in terms of accuracy and complexity. 
The component eigenmodes selected by the proposed RMI-based selection methods satisfied the accuracy requirements of the assembly ROM while significantly reducing the order of the original component and assembly models.
Furthermore, we compared the results obtained by the new RMI-based mode selection methods with a brute force selection method that computes the optimal selection of component eigenmodes (optimal in the sense of minimizing the number of component eigenmodes while still complying with the requested assembly ROM accuracy) and have showed that most of the RMI based mode selection methods obtained selections that were (close to) the optimal solutions, showing their effectiveness.

In conclusion, the proposed, RMI-based modular and assembly-accuracy-based component eigenmode selection approach, in combination with the robust-performance-based top-down approach, provides a practical and efficient framework for reducing the complexity of component models while maintaining the required accuracy of assembly ROM. 
As a result, in contrast to the standard industrial approach, it is no longer necessary to check the assembly accuracy after obtaining the assembly ROM.
Furthermore, the framework can be applied using any CMS method of choice.

\section{ACKNOWLEDGEMENTS}
This publication is part of the project Digital Twin with project number P18-03 of the research programme Perspectief which is (mainly) financed by the Dutch Research Council (NWO).

\end{document}